\documentclass[11pt]{article}
\usepackage[utf8]{inputenc} 
\usepackage{braket}
\usepackage{amsmath,latexsym}
\DeclareMathOperator{\arsinh}{arsinh}
\DeclareMathOperator{\artanh}{artanh}
\usepackage{dsfont}
\usepackage{physics}
\usepackage{indentfirst}
\usepackage{hhline}
\usepackage{amsfonts}
\usepackage[table]{xcolor}
\usepackage{graphicx}
\usepackage{pgfplots}
\usepackage{subfig}

\usepackage[style=numeric,backend=biber,sorting=none]{biblatex}
\usepackage{authblk}
\usepackage{todonotes}
\usepackage{array}
\usepackage{hyperref}
\hypersetup{
    colorlinks=true,
    linkcolor=blue,
    filecolor=magenta,      
    urlcolor=cyan,
}
\pgfplotsset{width=9cm,compat=1.9}
\graphicspath{ {images/} }
\usepackage{stackengine}

\title{Toward the Pole}

\author[1,2]{S. Alekseev}
\author[3,4]{A. Gorsky}
\author[1,2]{M. Litvinov}
\affil[1]{Institute for Theoretical and Experimental Physics, Moscow 117218, Russia}
\affil[2]{Department of Physics and Astronomy, Stony Brook University, Stony Brook, NY 11794-3800, USA}
\affil[3]{Institute for Information Transmission Problems of the Russian Academy of Sciences, Moscow 127051, Russia}
\affil[4]{Moscow Institute of Physics and Technology, Dolgoprudny 141700, Russia}
\begin{document}

\maketitle

\begin{abstract}
 We study the analytic structure of semiclassical conformal blocks, namely of the 1-point conformal block on the torus and of the 4-point conformal block on the sphere, as functions of the intermediate dimension. We interpret their discontinuities, which can be revealed with the use of a particular resummation procedure, holographically in terms of configurations of geodesics in AdS${}_3$. In other words, we study the behavior of the $\Omega$-deformed $\mathcal N=2$ SYM theory in the Nekrasov-Shatashvili limit near those singular points, where naively the W-boson with non-vanishing angular momentum becomes massless. Upon a proper resummation of instanton contributions, these singularities disappear, which is similar to the Seiberg-Witten solution in the undeformed case where there is no massless W-boson. It is shown that the order parameter undergoes a non-analytic behavior near positions of the poles. The jump of the order parameter is interpreted holographically in terms of geodesic networks.

\end{abstract}
\newpage

\tableofcontents
\newpage

\section{Introduction}

The Seiberg-Witten solution to the $\mathcal N=2$ SYM theory \cite{SW} provides an example of evaluation
of the spectrum of stable particles in the theory. In particular, it was demonstrated exactly
that the spectrum undergoes changes at some submanifolds in the moduli space \cite{Bilal}. In the simplest
case of the $SU(2)$ gauge group the W-boson turns out to be unstable near the origin of the moduli space
$a=0$, where $a$ is the vev of the adjoint scalar 
parametrizing the moduli space. The region of instability 
is enclosed by the curve of marginal stability, which involves two singular points 
where a monopole and a dyon become massless. There are nontrivial monodromies
of the spectrum around these points where the massless particle emerges.
Naively, the instanton contributions to the effective low-energy action  have poles
at the origin of the moduli space but upon resummation the naive singularity disappears.

The Seiberg-Witten solution has been derived explicitly upon introducing the
$\Omega$-background \cite{Nekrasov}, so that the powerful localization technique allowed
to evaluate the instanton contributions. Somewhat similar to the undeformed case,
naively there are points in the moduli space $a=n\epsilon_1$, where  W-bosons with non-vanishing angular momentum $n$ 
become massless. The deformation parameter  $\epsilon_1$ enters the partition function as the angular speed via the factor $\exp( -\epsilon_1 J)$, where $J$ is an operator of rotation in some plane in $\mathbb{R}^4$. There are poles in the instanton sums exactly at these points; however, with the use of
the approach from \cite{Beccaria}
it was shown in \cite{GorskyMS} that in the pure $SU(2)$ gauge theory the effective twisted superpotential can be resummed in such a way that naive poles recollect into branch cuts. Similar to the undeformed case,
the spectrum of the stable states is expected to be rearranged near the poles.

According to the AGT correspondence, the instanton part of the Nekrasov partition function in 4d is equal to a particular conformal block in 2d, whose type depends on matter content of the supersymmetric gauge theory \cite{AGT,2GaiDu}. In particular, the partition function of the $\mathcal N=2$ four-dimensional $SU(2)$ gauge theory without matter fields corresponds to the irregular conformal block in two-dimensional theory \cite{GaiottoIrreg}, and the effective twisted superpotential arising in the Nekrasov-Shatashvili limit \cite{NS} corresponds to the semiclassical irregular conformal block. In the same manner, partition functions of the $\mathcal N=2^{*}\ SU(2)$ theory with adjoint matter and of the $\mathcal N=2\ SU(2)$ theory with four matter fields in the fundamental representation correspond to the 1-point conformal block on a torus and to the 4-point conformal block on a sphere respectively.  

It is also useful to utilize the integrability trick.
It has been recognized that all low-energy information  concerning the $SU(2)$ and $SU(N)$ SYM theories  \cite{gkmmm,MN} is stored in $SU(2)$ and
$SU(N)$ classical closed Toda chains. Similarly, the $\mathcal{N}$=$2^*$ $SU(N)$ SYM with adjoint 
matter corresponds to the classical $SU(N)$ elliptic Calogero model \cite{dw}.
The idea behind integrability is quite simple. Namely, we add a proper probe defect into the theory and capture the information about the theory from the dynamics of the probe. The correspondence with the integrable systems becomes even more clear in the $\Omega$-deformed case when they get quantized \cite{NS}. The Schr{\"o}dinger equation for the wave function of the finite-dimensional Hamiltonian system is the same as that, which defines the semiclassical conformal block and which is obtained on the CFT side from the null-vector decoupling equation in the semiclassical limit, when one inserts the degenerate operator into a corresponding correlator (see for instance \cite{Nekrasov-defect}).

%%Depending on the four-dimensional theory, on the CFT side, one adds an additional degenerate operator into a correlator, which induces the second order differential equation of the Schr{\"o}dinger type [REF?]. In the semiclassical limit, the conformal block itself can be found by requiring a particular monodromy of the solution to this equation, which is the wave function of a corresponding finite-dimensional Hamiltonian system.

Namely, the problem of finding the semiclassical 1-point torus conformal block is equivalent to the spectral problem of the Lame equation \cite{Fateev}. From the point of view of conformal field theory, the Lame equation arises as the null-vector decoupling equation for the 2-point correlator on the torus \cite{Piatek_Torus}. In a proper limit the Lame equation turns into the Mathieu equation, whose solution defines the semiclassical irregular conformal block \cite{Jap,Piatek}. The spectrum of the Mathieu equation consists of a sequence of bands and gaps \cite{Dunne}, so the semiclassical irregular conformal block has an interesting analytic structure. The 4-point conformal block on the sphere can be calculated by solving the Heun equation and requiring a particular monodromy from its solutions \cite{Z2}. 

Recently, a significant progress has been made in identifying conformal blocks with different objects in AdS${}_3$. It has been shown that the global 4-point conformal block on the sphere corresponds to the so called geodesic Witten diagram – the diagram in a particular quantum field theory on the AdS${}_3$ background, where positions of vertices over which integration is performed are restricted to geodesics connecting points at the boundary \cite{Hijano_Geodesic}. The semiclassical 4-point conformal block in the heavy-light limit was identified with the action of the geodesic configuration in the geometry of the conical defect in AdS${}_3$ \cite{HijanoKraus}. Analogous results were also obtained for conformal blocks on the torus. The global 1-point conformal block on the torus was shown to correspond to the Witten diagram in thermal AdS${}_3$ \cite{Hijano_Torus}. In \cite{Alkalaev,Alkalaev_npt} the interpretation of the semiclassical $n$-point conformal block on the torus in terms of geodesics in thermal AdS${}_3$ was suggested.

We shall show very explicitly how the jumps of the superpotential/conformal block at the cuts are represented in terms of configurations of simplest geodesic networks. To this aim the special linearized limit of the semiclassical conformal block is considered. We will analyze
several types of conformal blocks, corresponding to geodesic networks in thermal AdS${}_3$ and in the background of the conical defect and of the BTZ black hole. In all the cases the corresponding rearrangement of the geodesic configuration will be identified. We will start, however, with the discussion of analytic properties of the semiclassical irregular conformal block, which does not seem to have any simple holographic interpretation, but which is crucial for the whole topic. 

The paper is organized as follows.
The main text consists of 3 parts, each of which is devoted to different types of classical conformal blocks. In Section \ref{irreg} the resummation procedure of the semiclassical irregular conformal block and its consequences in CFT${}_2$ are discussed.  In Section \ref{torus} we analyze the analytic structure of the linearized semiclassical torus conformal block with the use of an analogous resummation procedure and of its holographic interpretation. In Section \ref{4pt} we analyze the analytic structure of the semiclassical heavy-light 4-point conformal block on the sphere with  the use  of its  holographic  interpretation. In Conclusion we formulate the main results of the paper and mention a bunch of open questions. The relation between geodesics in the conical defect and in the BTZ black hole backgrounds and also the network of geodesics at a constant angle in the BTZ background are discussed in Appendices.  

\section{Irregular conformal block}\label{irreg}
In this section we briefly review the definition of the irregular conformal block, methods of calculating it, and how it is related to the spectrum of the Mathieu equation in the semiclassical limit. Then we discuss a double series representation of the semiclassical irregular conformal block, which reveals its discontinuities at some discrete values of the conformal dimension.
\par The irregular conformal block is defined as the norm of the Gaiotto state $\ket{\Delta,\Lambda^2}$ \cite{GaiottoIrreg}:  
\begin{equation}
\label{eq:irreg_def}
\mathcal F_{\operatorname{irr}}(\Delta,c|\Lambda^2)=\Braket{\Delta,\Lambda^2|\Delta,\Lambda^2}.
\end{equation}
By definition, the Gaiotto vector $\ket{\Delta,\Lambda^2}$ is a linear combination of vectors belonging to the Verma module of some primary vector $\ket{\Delta}$, which satisfies to the following conditions:
\begin{equation}
\label{eq:gaiotto_eqns}
\begin{split}
& L_1 \ket{\Delta,\Lambda^2}=\Lambda^2\ket{\Delta,\Lambda^2},\\ 
& L_n \ket{\Delta,\Lambda^2}=0,\ n\ge 2.
\end{split}
\end{equation}
One can solve these equations order by order in $\Lambda^2$, looking for a solution in the form $\ket{\Delta,\Lambda^2}=\sum_{n=0}^{\infty}\Lambda^{2n}\ket{v_n},$ where $\ket{v_0}=\ket{\Delta}$, $L_0 \ket{v_n}=(\Delta+n)\ket{v_n}$. It is also known that the irregular conformal block can be obtained from the 4-point conformal block on the sphere ${\mathcal{F}}_{\operatorname{4pt}}(\Delta,\Delta_i,c|z)$ in the limit, when some external dimensions tend to infinity, while the coordinate $z$ tends to $0$ \cite{MMM}:
\begin{equation}
\label{eq:lim}
\begin{split}
&\epsilon_2,\epsilon_3 \to -\infty,\ z \to 0,\\
&\epsilon_2\epsilon_3 z\equiv\Lambda^4= \operatorname{const},
\end{split}
\end{equation}
where $\epsilon_{i}=b^2 \Delta_{i}\big|_{b\to 0}$ are external dimension in the semiclassical limit, and we use the standard parametrization of the central charge $c\equiv1+6(b^{-1}+b)^2\equiv1+6Q^2$. So, the following equality, relating these two conformal blocks, holds:
\begin{equation}
\begin{split}
\label{eq:exp_fromOPE}
&\mathcal{F}_{\operatorname{irr}}(\Delta,c|b^{-2}\Lambda^2)=\lim_{\substack{\epsilon_{2,3}\to -\infty \\ z\to 0}} \mathcal F_{\operatorname{4pt}}(\Delta,\Delta_i,c|z)=\sum_{N\ge 0} b^{-4N} \big(\mathcal N^{-1}_{\Delta}\big)_{\{1^N;1^N\}}\, \Lambda^{4N},
\end{split}
\end{equation}
where $\big(N^{-1}_{\Delta}\big)_{\{1^N;1^N\}}$ are the elements of the inverse Gram matrix of basis vectors in the Verma module, built over the primary vector $\ket{\Delta}$:
\begin{equation}
(\mathcal{N}_\Delta)_{\,Y'Y}\equiv\bra{\Delta}L_{Y'}L_{-Y}\ket{\Delta}.
\end{equation}
From Zamolodchikov's recursion formula for the 4-point conformal block on the sphere \cite{Z1,Z2,Perlmutter}, taking the limit of large conformal dimensions, one can also obtain the recursion formula for the irregular conformal block \cite{Poghossian}, which is convenient for actual computations of conformal block coefficients:
\begin{equation}
\label{eq:recurs_H}
\mathcal F_K(\Delta,c)=\sum_{mn+N=K}\frac{ A_{mn}(c)}{\Delta-\Delta_{mn}}\ \mathcal F_N(\Delta_{mn}+mn,c),\ \mathcal F_0=1,
\end{equation}
where the notation $\mathcal F_{N}(\Delta,c)\equiv \big(\mathcal N^{-1}_{\Delta}\big)_{\{1^N;1^N\}}$ was introduced for elements of the inverse Gram matrix, and $\Delta_{mn}$ are the degenerate dimensions, which are given explicitly below. The coefficients $A_{mn}(c)$ are defined as follows: 
\begin{equation}
\label{eq:amn}
A_{mn}(c)=\frac{(-1)^{m+n}}{2}\prod_{p,q}\frac{1}{pb^{-1}+qb},    
\end{equation}
where the numbers $p$ and $q$ in the product run over the following sets of integers: $p=-m+1,-m+2,\ldots,m;\ q=-n+1,-n+2,\ldots,n$, excluding the pairs $(p,q)=(0,0)$ and $(m,n)$. 
\par The problem of calculating the semiclassical irregular conformal block
\begin{equation}
\label{eq:quasi_irr} f_{\operatorname{irr}}(\epsilon|\Lambda^2)\equiv \lim_{b\to 0}b^2\log\mathcal{F}_{\operatorname{irr}}(\Delta,c|b^{-2}\Lambda^2)
\end{equation}
(the argument rescaling $\Lambda^2 \to b^{-2}\Lambda^2$ is necessary for the conformal block to exponentiate in the semiclassical limit) is equivalent to the problem of finding the spectrum of the Mathieu equation, describing a particle in the periodic potential,
\begin{equation}
\label{eq:mathieu}
-\psi''(z)+2\Lambda^2\cos(2z)\psi(z)=u(\nu,\Lambda^2)\psi(z)  
\end{equation}
for quasiperiodic wave functions $\psi(z+\pi)=e^{\pi i \nu}\psi(z)$ \cite{Jap,Piatek}. The quasimomentum $\nu$ of the particle is related to the conformal dimension in the semiclassical limit:
\begin{equation}
\label{eq:clas_parametr_mathieu}
\epsilon\equiv\frac{1-\nu^2}{4}\equiv    b^2\Delta\big|_{b\to 0},    
\end{equation}
and the energy of the particle $u(\nu,\Lambda^2)$ is directly related to the semiclassical irregular conformal block:
\begin{equation}
\label{eq:matone}
u(\nu,\Lambda^2)=-\frac{\Lambda}{4}\frac{\partial}{\partial\Lambda} f_{\operatorname{irr}}(\nu,\Lambda^2)+\operatorname{const}.   
\end{equation}
The particle energy also has a natural interpretation in CFT${}_2$ as the expectation value of the operator $L_0$ over the Gaiotto state, or rather as the deviation of this expectation value from the value $\Delta$, as the following equation holds:
\begin{equation}
\label{eq:whatisu}
-\frac{\Lambda}{4}\frac{\partial}{\partial\Lambda} f_{\operatorname{irr}}(\nu,\Lambda^2)=b^2 \frac{\Braket{\Delta,\Lambda^2|L_0-\Delta|\Delta,\Lambda^2}}{\braket{\Delta,\Lambda^2}{\Delta,\Lambda^2}}\bigg|_{b\to 0},
\end{equation}
which follows from the equality:
\begin{equation}
\frac{\Lambda}{2} \frac{\partial}{\partial \Lambda}\ket{\Delta,\Lambda^2}=(L_0-\Delta)\ket{\Delta,\Lambda^2}.
\end{equation}
The first few terms in the expansion of the semiclassical conformal block in the parameter $q\equiv\Lambda^2$, which can be obtained either from the definitions (\ref{eq:irreg_def}) and (\ref{eq:quasi_irr}) or from solving the Mathieu equation perturbatively in the potential amplitude, have the following form:
\begin{equation}
\label{eq:irreg_block_exp}
\begin{split}
&f_{\operatorname{irr}}(\epsilon,q)=\frac{1}{2\epsilon}q^2- \frac{3-5\epsilon}{16\epsilon^3(3+4\epsilon)}q^4+\frac{12 - 38\epsilon + 18\epsilon^2}{96 \epsilon^5 (2 + \epsilon) (3 + 4 \epsilon)}q^6+O(q^8)=\\&=-\frac{2}{\nu^2-1} q^2-\frac{5\nu^2+7}{(\nu^2-1)^3(\nu^2-4)}q^4-\frac{16(9\nu^4+58\nu^2+29)}{3(\nu^2-1)^5(\nu^2-4)(\nu^2-9)}q^6+O(q^8). 
\end{split}
\end{equation}
With the use of the formula (\ref{eq:matone}) an analogous expansion, singular at integer values of the parameter $\nu$, can be also obtained for the energy $u(\nu,\Lambda^2)$. From the general theory of the Mathieu equation it is known, however, that for every $\nu \in \mathbb{Z}$ the equation (\ref{eq:mathieu}) has solutions, corresponding to top and bottom edges of gaps and bands in the spectrum \cite{Dunne}. A certain procedure of resummation of the irregular conformal block (\ref{eq:irreg_block_exp}) allows one to obtain energy values at the edges of bands/gaps and gives new information about the analytic structure of the conformal block \cite{GorskyMS,Beccaria,AlekseevLitvinov}. To obtain a good expansion of the irregular conformal block near integer values of $\nu$, one reorganizes its expansion in $q$ as a double series:
\begin{equation}
\label{eq:resum}
f_{\operatorname{irr}}(\nu,q)=\sum_{n=1}^{\infty}\sum_{k=1}^{\infty}\left[g^{(n)}_{\, k}\bigg(\frac{q^n}{n-\nu}\bigg)+g^{(n)}_{\, k}\bigg(\frac{q^n}{n+\nu}\bigg)\right]q^{2k-2+n}, 
\end{equation}
where functions $g^{(n)}_{\, k}$ have the following form: 
\begin{equation}
\label{eq:gnk}
\begin{split}
&g^{(n)}_{\, 1}\left(z\right)=\frac{1-\log2+\log\left(1+\sqrt{1+4z^2/\zeta_n^2}\right)-\sqrt{1+4z^2/\zeta_n^2}}{z},\\
&g^{(n)}_{\, k} (z)=\frac{\left(1+4z^2/\zeta_n^2\right)^{\frac{5}{2}-k}Q_{2k-3}(z^2)+P_{k-1}(z^2)}{z^{2k-1}},\ k>1,
\end{split}
\end{equation}
where $\zeta_n=n!(n-1)!$ and $Q_m(z),P_m(z)$ are some polynomials of degree $m$, whose form depends on $n$ and $k$. The explicit form of the first few functions $g^{(n)}_{\, k}$ with $k>1$ is given in the work \cite{GorskyMS}. Square roots $\sqrt{1+\frac{4q^{2n}/\zeta_n^2}{(n\pm\nu)^2}}$, entering functions $g^{(n)}_{\, k}$ with $k\in \mathbb{N}$, have branching points at $\nu=n,\, n\pm 2iq^n/\zeta_n$. Every function $g^{(n)}_{\, 1}, n\in \mathbb{N}$ contains also a logarithm with additional branching points $\nu=n,\, \infty$. On the complex plane $\nu \in \mathbb{C}$ with branch cuts, extending from the points $n+2iq^n/\zeta_n,\, n\in \mathbb{N}$ to infinity and passing through the points $\nu=n,\, n-2iq^n/\zeta_n$ (Fig. \ref{fig:branch_cuts}), one can choose branches of all the multivalued functions entering the expression (\ref{eq:resum}) in such a way that all the square roots will be positive at real $\nu$, and all the logarithms will be real. On this branch the expansion of the double series (\ref{eq:resum}) in $q$ reproduces the expansion of the conformal block (\ref{eq:irreg_block_exp}). It is not difficult to see that this double series has finite and different limits at $\nu\to n\pm 0$. For instance, at $\nu \to 1 \pm 0$ we obtain the following limits of the irregular conformal block (the two series below differ only in signs in front of odd powers of $q$):
\begin{figure}
\centering
\includegraphics[scale=0.28]{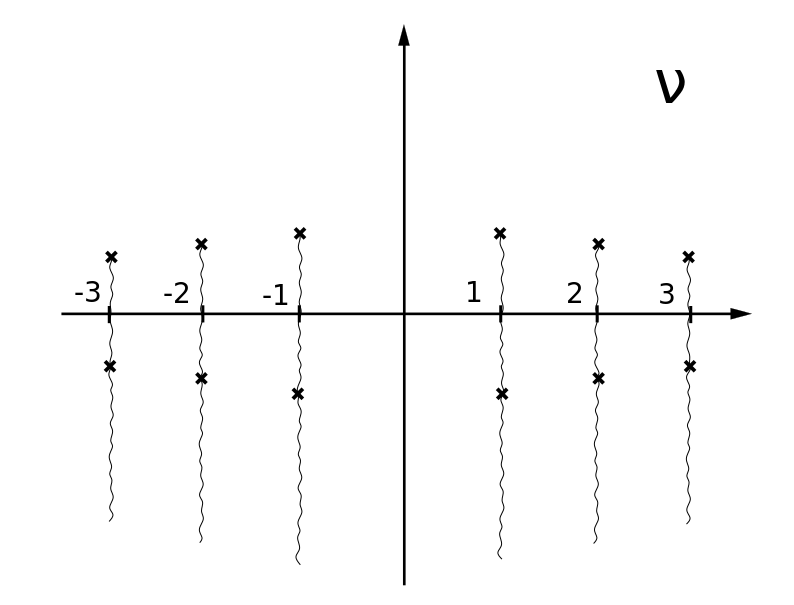}
\caption{Branching points of the functions $g^{(n)}_{\, k}$, entering the double series of the semiclassical irregular conformal block $f_{\operatorname{irr}}(\nu,q)$, on the plane of intermediate dimensions $\nu$. Branch cuts are depicted as wavy lines.}
\label{fig:branch_cuts}
\end{figure}
\begin{equation}
\begin{split}
&f_{\operatorname{irr}}(\nu,q)\big|_{\nu\to1+0}=2q-\frac{q^2}{2}-\frac{q^3}{6}-\frac{q^4}{48}+O(q^5),\\
&f_{\operatorname{irr}}(\nu,q)\big|_{\nu\to1-0}=-2q-\frac{q^2}{2}+\frac{q^3}{6}-\frac{q^4}{48}+O(q^5).
\end{split}
\end{equation}
It means that the semiclassical irregular conformal block changes discontinuously at integer values of $\nu$:
\begin{equation}
\label{eq:ireg_discontinuity}
\lim_{b\to 0}b^2 \log \Braket{\Delta_n^-,b^{-2}\Lambda^2|\Delta_n^-,b^{-2}\Lambda^2}\ne \lim_{b\to 0}b^2 \log \Braket{\Delta_n^+,b^{-2}\Lambda^2|\Delta_n^+,b^{-2}\Lambda^2},  
\end{equation}
where conformal dimensions $\Delta_n^{\pm}$ differ only in the sign of a small additive constant:
\begin{equation}
b^2\Delta_n^{\pm}\big|_{b\to 0}=\frac{1-n^2}{4}\mp 0.  
\end{equation}
What happens to the irregular conformal block when the parameter $\nu$ is exactly integer? First, the value of $\nu$ is equal to the integer number $n$ for degenerate dimensions $\Delta_{nm}$:
\begin{equation}
\Delta_{nm}=\frac{Q^2}{4}-\frac{1}{4}\left(nb^{-1}+mb\right)^2=\frac{1}{4}\left[b^{-2}(1-n^2)+2(1-nm)+b^2(1-m^2)\right],
\end{equation}
but Gaiotto states $\ket{\Delta,\Lambda^2}$ with such dimensions, satisfying conditions (\ref{eq:gaiotto_eqns}), do not exist. However, there is an infinite set of conformal dimensions with integer $\nu$, which do not define degenerate states:
\begin{equation}
\label{eq:baddim}
\Delta=b^{-2}\,\frac{1-n^2}{4}+\delta(b)
\end{equation}
where $\delta(b)=O(1)$ at $b\to 0$ and $\delta(b)\ne\frac{1}{4}\left[(1-nm)+b^2(1-m^2)\right]$. It appears that irregular conformal blocks with such dimensions do not exponentiate in the semiclassical limit. To demonstrate it, let us give the general form of the perturbative expansion of the irregular conformal block:
\begin{equation}
\label{eq:gaiotto_expansion}
\braket{\Delta,b^{-2}\Lambda^2}{\Delta,b^{-2}\Lambda^2}=1+\frac{ A_{11}}{\Delta-\Delta_{11}}\left(\frac{\Lambda}{b}\right)^4+\bigg[\frac{ A_{12}}{\Delta-\Delta_{12}}+\frac{ A_{21}}{\Delta-\Delta_{21}}+\frac{\left(A_{11}\right)^{ 2}}{\Delta-\Delta_{11}}\bigg]\left(\frac{\Lambda}{b}\right)^8+O\left(\Lambda^{12}/b^{12}\right),
\end{equation}
where coefficients $A_{nm}$ are defined in the formula (\ref{eq:amn}). For conformal dimensions having the form  (\ref{eq:clas_parametr_mathieu}) in the semiclassical limit the following equality holds:
\begin{equation}
\Delta-\Delta_{nm}=b^{-2}\,\frac{\nu^2-n^2}{4}+O(1),    
\end{equation}
so the behaviour of singularities in the conformal block expansion (\ref{eq:gaiotto_expansion}) changes significantly for $\nu \in \mathbb{Z}$. For instance, when $\nu=1$ and $\delta(b)=\delta_0=\operatorname{const}$, one obtains:
\begin{equation}
\log\braket{\Delta,b^{-2}\Lambda^2}{\Delta,b^{-2}\Lambda^2}=\left[-\frac{b^{-4}}{2\delta_0}+O(1)\right]\Lambda^4+\left[\frac{b^{-8}}{4\delta_0^2(2+4\delta_0)}+O(b^{-6})\right]\Lambda^8+O(\Lambda^{12}).
\end{equation}
For comparison, the irregular conformal block at non-integer $\nu$ has the following structure:
\begin{equation}
\label{eq:energy_discontinuity}
\begin{split}
&\log\braket{\Delta,b^{-2}\Lambda^2}{\Delta,b^{-2}\Lambda^2}=\left[-\frac{2}{\nu^2-1}b^{-2}+O(1)\right]\ \Lambda^4+\left[-\frac{5\nu^2+7}{(\nu^2-1)^3(\nu^2-4)}b^{-2}+O(1)\right]\Lambda^8+\\
&+ O(\Lambda^{12})=b^{-2}f_{\operatorname{irr}}(\nu,\Lambda^2)+O(b^{0}).
\end{split}
\end{equation} 
According to the relation (\ref{eq:matone}), differentiating the resummed conformal block (\ref{eq:resum}) with respect to $\log\Lambda$ gives the representation of the energy $u(\nu,q)$ as a double series. After differentiation, the logarithms contained in functions $g^{(n)}_{\, 1}$ disappear, so branching points of the function $u(\nu,q)$ are defined only by square roots $\sqrt{1+\frac{4q^{2n}/\zeta_n^2}{(n\pm\nu)^2}}$. The double series for the energy $u(\nu,q)$ has again different limits at $\nu\to n\pm 0$, reproducing known values of the energy at the bottom of the $(n+1)$th band and at the top of the $n$th band respectively \cite{GorskyMS}. From the point of view of CFT${}_2$, these different limits mean that the expectation value of the operator $L_0$ over the Gaiotto state changes discontinuously at $\nu \in \mathbb{Z}$:
\begin{equation}
b^2\,\frac{\Braket{\Delta_n^{+},\Lambda^2|L_0|\Delta_n^{+},\Lambda^2}}{\braket{\Delta_n^{+},\Lambda^2}{\Delta_n^{+},\Lambda^2}}\bigg|_{b\to 0}\ne b^2\,\frac{\Braket{\Delta_n^{-},\Lambda^2|L_0|\Delta_n^{-},\Lambda^2}}{\braket{\Delta_n^{-},\Lambda^2}{\Delta_n^{-},\Lambda^2}}\bigg|_{b\to 0}.    
\end{equation}
\section{One-point torus conformal block}\label{torus}
\subsection{Linearized limit of semiclassical torus conformal block}
In this section we define the linearized 1-point semiclassical conformal block on the torus, which will later appear from the holographic point of view, and show that it can be resummed in the same way as the semiclassical irregular conformal block, which, in particular, reveals its discontinuity at some value of the semiclassical intermediate dimension. 
\par The 1-point conformal block on the torus is defined as follows \cite{Hijano_Torus,Fateev}:
\begin{equation}
\tilde{\mathcal{F}}_{\operatorname{tor}}(\Delta_p,\Delta,c|q)=\Tr\left[P_{\Delta_p}q^{L_0-c/24}\phi_\Delta(z)\right],
\end{equation}
where $P_{\Delta_p}$ is a projector on the Verma module of the primary vector $\ket{\Delta_p}$, $\phi_\Delta(z)$ is a primary field having the dimension $\Delta$, and $q\equiv e^{2\pi i\tau}$, where $\tau$ is a modular parameter of the torus. As before, we consider the redefined conformal block  $\mathcal{F}_{\operatorname{tor}}(\Delta_p,\Delta,c|q)=q^{-\Delta_p+c/24} \tilde{\mathcal{F}}_{\operatorname{tor}}(\Delta_p,\Delta,c|q)$, such that $\mathcal{F}_{\operatorname{tor}}(\Delta_p,\Delta,c|0)=1$. The first few terms of the expansion of the semiclassical torus conformal block 
\begin{equation}
f_{\operatorname{tor}}(\epsilon_p,\epsilon|q)\equiv \lim_{b\to 0}b^2\log\mathcal{F}_{\operatorname{tor}}(\Delta_p,\Delta,c|q)=\sum\limits_{n=1}^{\infty} f_n(\epsilon_p,\epsilon)q^{n}
\end{equation}
in the parameter $q$ have the following form \cite{Piatek_Torus}:
\begin{equation}
\begin{split}
\label{eq:toricblock_coeff}
&f_1(\epsilon_p,\epsilon)=-\frac{\epsilon^2}{1-\nu} - \frac{\epsilon^2}{1+\nu},\\
&f_2(\epsilon_p,\epsilon)= -\frac{4 \epsilon^2 + 4 \epsilon^3 + \epsilon^4}{4 (2 - \nu)} + \frac{\epsilon^4}{2 (1 - \nu)^3} + \frac{
 -2 \epsilon^2 + 4 \epsilon^3 + \epsilon^4}{2 (1 - \nu)}+\\
 &+ \frac{-2 \epsilon^2 + 4 \epsilon^3 + \epsilon^4}{
 2 (1 + \nu)}+ \frac{\epsilon^4}{
 2 (1 + \nu)^3} - \frac{4 \epsilon^2 + 4 \epsilon^3 + \epsilon^4}{4 (2 + \nu)},
\end{split}
\end{equation}
where the following notation for conformal dimensions in the semiclassical limit was introduced:
\begin{equation}
\label{eq:dim_torus}
\epsilon\equiv b^2\Delta\big|_{b\to 0},\ \epsilon_p\equiv\frac{1-\nu^2}{4}\equiv    b^2\Delta_p\big|_{b\to 0}.
\end{equation}
For comparison, let us also represent the first expansion coefficients of the semiclassical irregular conformal block   (\ref{eq:irreg_block_exp}) in the same form:
\begin{equation}
\begin{split}
&f^{(\operatorname{irr})}_1(\nu)=-\frac{1}{1-\nu} - \frac{1}{1+\nu},\\
&f^{(\operatorname{irr})}_2(\nu)=- \frac{1}{4 (2 - \nu)} + \frac{1}{2 (1 - \nu)^3} + \frac{1}{2 (1 - \nu)} + \frac{1}{
 2 (1 + \nu)}+ \frac{1}{
 2 (1 + \nu)^3}  - \frac{1}{4 (2 + \nu)}.
\end{split}
\end{equation}
So, one sees that the expansion coefficients of the conformal block on the torus and of the irregular conformal block are related to each other as follows:
\begin{equation}
f^{(\operatorname{irr})}_n(\nu)=\lim_{\epsilon\to \infty} \epsilon^{-2n}f_n(\nu,\epsilon).
\end{equation}
Therefore, to obtain the semiclassical irregular conformal block from the semicalssical conformal block on the torus, one has to take the limit $\epsilon\to \infty,\ q\to 0,\ \epsilon^2 q \to \operatorname{const}$ in the latter and introduce the redefinition $\epsilon^2 q=\Lambda^4$.  By definition, the linearized semiclassical conformal block on the torus is obtained from the full semiclassical torus conformal block in the limit $\epsilon_p \to 0,\ \epsilon\to 0,$ while $\delta\equiv \epsilon/\epsilon_p \to \operatorname{const}$ \cite{Alkalaev}:
\begin{equation}
\label{eq:linear_toric}
f_{\, \operatorname{tor}}^{\, \operatorname{lin}}(\epsilon_p,\epsilon|q)=\epsilon_p \lim_{\substack{\epsilon_p\to0 \\ \epsilon\to 0}} \epsilon_p^{-1} f_{\, \operatorname{tor}}(\epsilon_p,\epsilon|q).  
\end{equation}
Comparing the expansion in $q$ of the linearized conformal block with the expansion of the full torus conformal block (\ref{eq:toricblock_coeff}), one can see, which part of the full semiclassical torus conformal block survives in this limit:
\begin{equation}
\label{eq:linearblock_coeff}
\begin{split}
&f_{\,1}^{\, \operatorname{lin}}( \epsilon_p,\epsilon)=- \epsilon_p\, \frac{\delta^2}{2}=-\frac{\epsilon^2}{1-\nu},\\
&f_{\,2}^{\, \operatorname{lin}}(\epsilon_p,\epsilon)=\frac{ \epsilon_p}{16}\left(-8\delta^2+\delta^4\right)=\frac{\epsilon^4}{2 (1 - \nu)^3} - \frac{
\epsilon^2}{(1 - \nu)},\\
&f_{\,3}^{\, \operatorname{lin}}( \epsilon_p,\epsilon)=\frac{ \epsilon_p}{48} \left(-24 \delta^2 + 8 \delta^4 - \delta^6\right)=-\frac{2 \epsilon^6}{3 (1-\nu)^5} + \frac{4 \epsilon^4}{3 (1-\nu)^3}- \frac{\epsilon^2}{1 - \nu}.
\end{split}
\end{equation}
In the formulae above the equality $\epsilon_p=(1-\nu)/2$ has been used, which is obtained from (\ref{eq:dim_torus}) in the limit $\nu\to 1$. It can also be seen from these formulas that the most singular (when $\nu \to 1$) terms at each order in the variable $q$ add up to the familiar function $g^{(1)}_{\, 1}$ from the formula (\ref{eq:gnk}), which has branching points and different limits when $\nu\to 1\pm 0$:
\begin{equation}
\label{eq:torus_g11}
\begin{split}
&-\frac{\epsilon^2}{1-\nu}\,q^2+\frac{\epsilon^4}{2 (1 - \nu)^3}\,q^4 -\frac{2 \epsilon^6}{3 (1-\nu)^5}\,q^6+...=\\
&=\left[-\frac{\epsilon q}{1-\nu}+\frac{(\epsilon q)^3}{2 (1 - \nu)^3} -\frac{2 (\epsilon q)^5}{3 (1-\nu)^5}+...\right]\epsilon q=g^{(1)}_1\left(\frac{\epsilon q}{1-\nu}\right)\epsilon q.
\end{split}
\end{equation}
The linearized semiclassical torus conformal block can be resummed in the same way as the semiclassical irregular conformal block was resummed in (\ref{eq:resum}):
\begin{equation}
\label{eq:torus_resum}
f_{\, \operatorname{tor}}^{\, \operatorname{lin}}\left(\epsilon_p=\frac{1-\nu}{2},\epsilon\bigg|q\right)=\epsilon\sum_{k\ge1}g^{(1)}_{\, k}\left(\frac{\epsilon q^{1/2}}{1-\nu}\right)q^{(2k-1)/2}.  
\end{equation}
The first two functions $g^{(1)}_{\, k},\ k>1$ have the following form:
\begin{equation}
\begin{split}
& g^{(1)}_{\, 2}(z)=\frac{-1+(1-2z^2)\sqrt{1+4z^2}}{6z^3},\\
& g^{(1)}_{\, 3}(z)=\frac{-1+(1+4z^2)^{-1/2}\left(1+2z^2-2z^4-16z^6\right)}{20z^5}.
\end{split}
\end{equation}
The series obtained has different limits when $\nu\to1\pm 0$, which differ only in the sign. Note also that at $\nu\to1\pm 0$ the conformal block is expanded only in odd powers of $\sqrt{q}$. In the next section the linearized semiclassical conformal block on the torus will be calculated analytically.
\subsection{Holographic interpretation of semiclassical torus conformal block}
The linearized semiclassical 1-point conformal block on the torus has a simple holographic interpretation as the length of the tadpole graph embedded in thermal AdS${}_3$ (Fig. \ref{fig:tadpole}), which is a torus topologically. This interpretation was suggested in the work \cite{Alkalaev}, but the non-perturbative answer for the linearized torus conformal block was not given there. We close this gap and also generalize the analysis of geodesics to the case of negative intermediate dimensions, which corresponds to the case $\nu>1$ and which allows us to obtain the holographic interpretation of the discontinuity mentioned above. We are looking for the extrema of the following action:
\begin{equation}
\label{eq:total_action}
S=\epsilon_p S_{\operatorname{loop}}+\epsilon S_{\operatorname{leg}},     
\end{equation}
where $S_{\operatorname{loop}}$ is the length of the loop, and $S_{\operatorname{leg}}$ is the length of the leg, which is attached at one end to the boundary of AdS${}_3$. Thus, the semiclassical conformal dimension $\epsilon_p$ in the torus conformal block is identified with the mass of a particle in the loop, while the external dimension $\epsilon$ is identified with the mass of a particle, propagating along the leg. The modular parameter of the torus in CFT${}_2$ is related to the length of the thermal cycle in AdS${}_3$, the metric on which has the following form:
\begin{figure}
\centering
\includegraphics[scale=0.25]{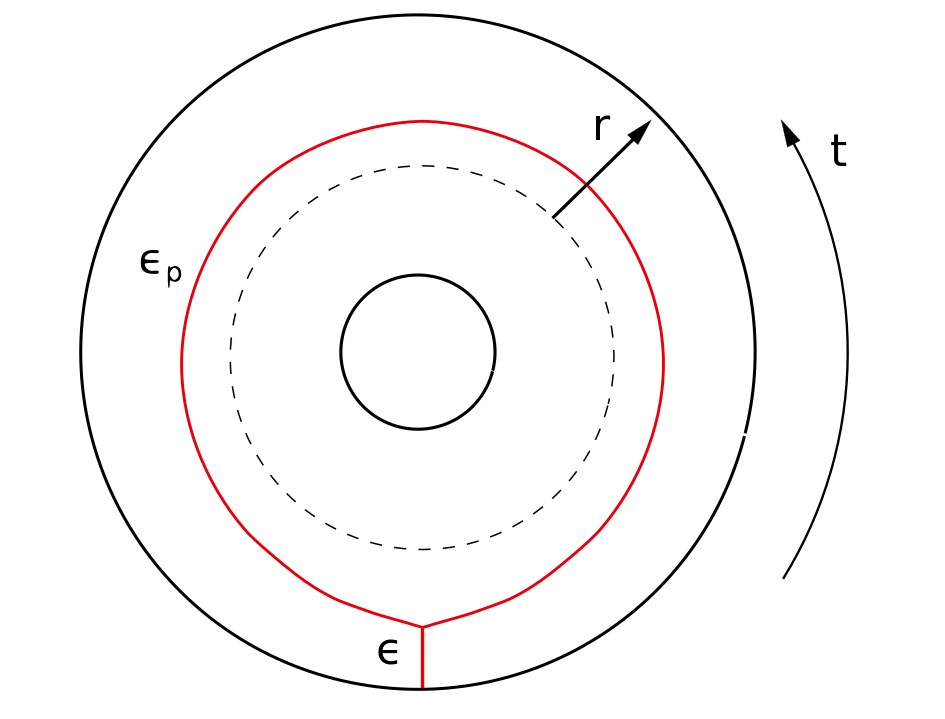}
\caption{The configuration of geodesics in thermal AdS${}_3$, which is holographically dual to the linearized semiclassical 1-point conformal block on the torus. The intersection of the solid torus with the plane, in which the circle $r=0$ (represented by a dotted line) lies, is shown.}
\label{fig:tadpole}
\end{figure}
\begin{equation}
ds^2=-\tau^2\left(1+\frac{r^2}{l^2}\right)dt^2+\left(1+\frac{r^2}{l^2}\right)^{-1}dr^2+r^2d\varphi^2, 
\end{equation}
where it is supposed that the modular parameter is purely imaginary $\tau=i|\tau|$, and $t\sim t+2\pi$, $0\le r<\infty$, $\varphi\sim \varphi+2\pi$. It will be shown later that up to a constant not depending on $q\equiv e^{2\pi i\tau}$ the following equation holds:
\begin{equation}
f_{\, \operatorname{tor}}^{\, \operatorname{lin}}(\epsilon_p,\epsilon|q)=S^{(\operatorname{extr})}(\epsilon_p,\epsilon|q),    
\end{equation}
where in the r.h.s. we have the action (\ref{eq:total_action}), calculated on the extremal configuration. Later it will be shown that the r.h.s. only exists when $|\epsilon/\epsilon_p|<2$. Only those geodesic configurations are considered, which belong to the slice of the torus, shown in Fig. \ref{fig:tadpole}. For the correct formulation of the variational problem, namely for eliminating boundary terms, appearing due to variation of the action (\ref{eq:total_action}), it is necessary to require compliance with the law of momentum conservation in the vertex (the subscripts are defined in Fig. \ref{fig:vertices}):
\begin{equation}
\label{eq:momentum_conserv_torus}
\epsilon_p\left(\dot x^{\mu}_2-\dot x^{\mu}_1\right)+\epsilon \dot x^{\mu}_0=0,  
\end{equation}
where the derivatives are taken with respect to the proper length along each of the geodesics. We parametrize the loop in such a way that increasing the proper length along the loop corresponds to the growth of time $t$. The leg parametrization is chosen such that the velocity at each its point is directed toward the vertex. As usual, instead of solving geodesic equations, we use the momentum conservation law along geodesics and the equation $u^2=1$:
\begin{equation}
\begin{split}
\label{eq:geodes}
&p_t=g_{tt}\frac{dt}{ds}=\operatorname{const}>0,\\  
& g_{tt}
\left(\frac{dt}{ds}\right)^2+ g_{rr}
\left(\frac{dr}{ds}\right)^2=1.
\end{split}
\end{equation}
We suppose that the time coordinate $t$ changes from $-\pi$ to $\pi$, so the boundary condition for the loop can be chosen in the following form:
\begin{equation}
\label{eq:bound_cond_loop}
r(-\pi)=r(\pi)=\rho,
\end{equation}
where $\rho$ is the radial coordinate of the vertex. As the equality $\dot t_2=\dot t_1$ holds for the loop, it follows from the momentum conservation law in the vertex (\ref{eq:momentum_conserv_torus}), that for the leg $\dot t_0=0$. Radial components of the velocity vector are related to each other as follows in the vertex:
\begin{equation}
\label{eq:momentum}
2\dot r_2=-\frac{\epsilon}{\epsilon_p}\,\dot r_0.  
\end{equation}
Therefore, for each sign of the intermediate dimension $\epsilon_p$ (the external dimension $\epsilon$ is supposed to be positive everywhere in the text) there are two options for the behaviour of geodesics in the vicinity of the vertex (Fig. \ref{fig:vertices}):
\begin{figure}
\begin{tabular}{cccc}
  \includegraphics[width=40mm]{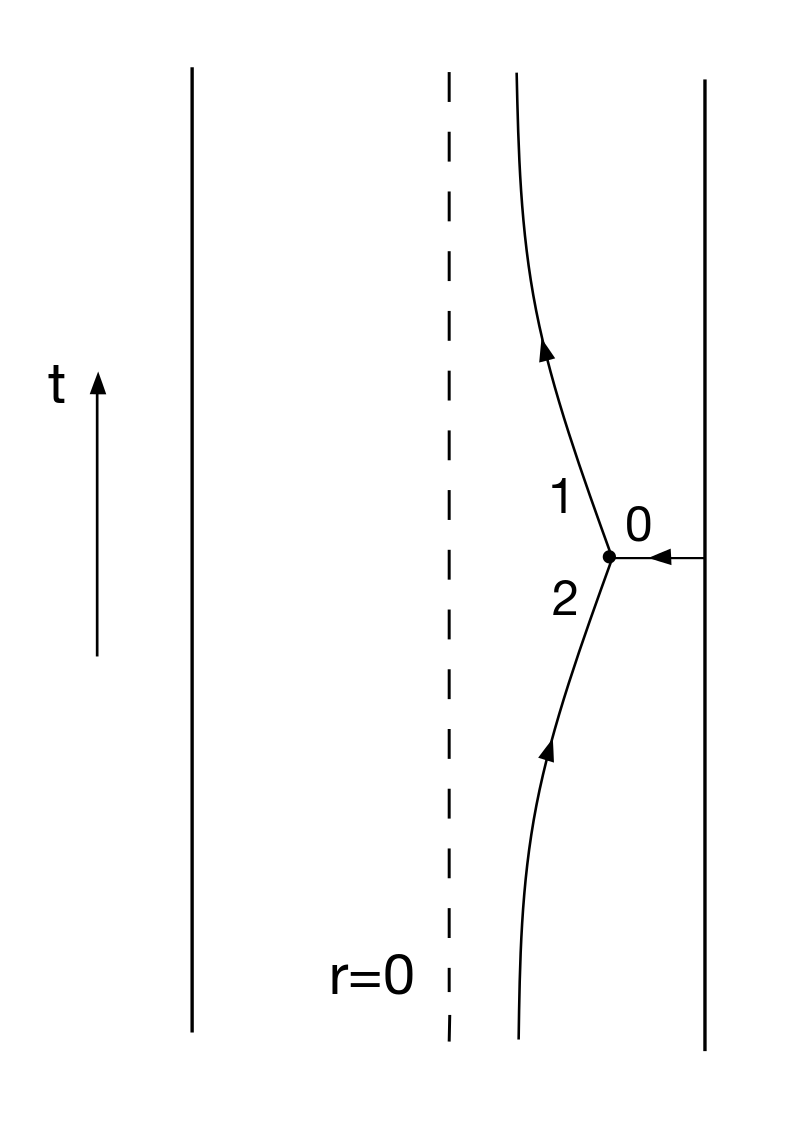} &   \includegraphics[width=40mm]{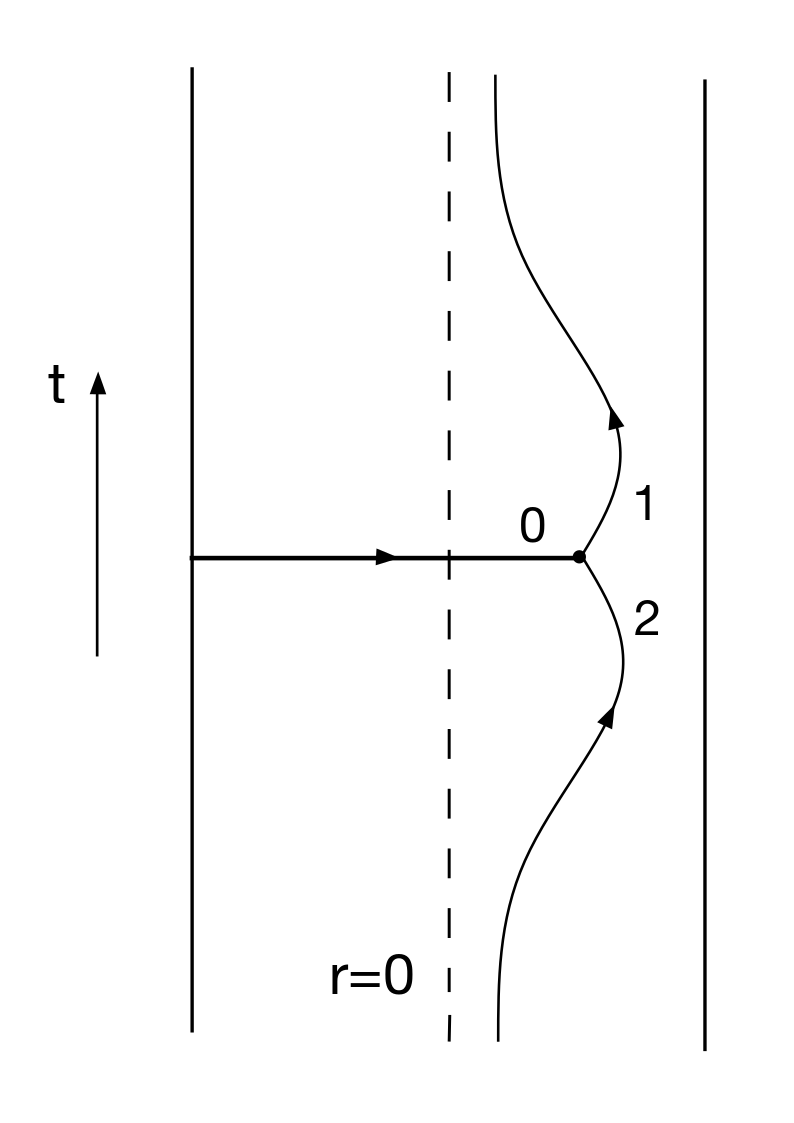} &   \includegraphics[width=40mm]{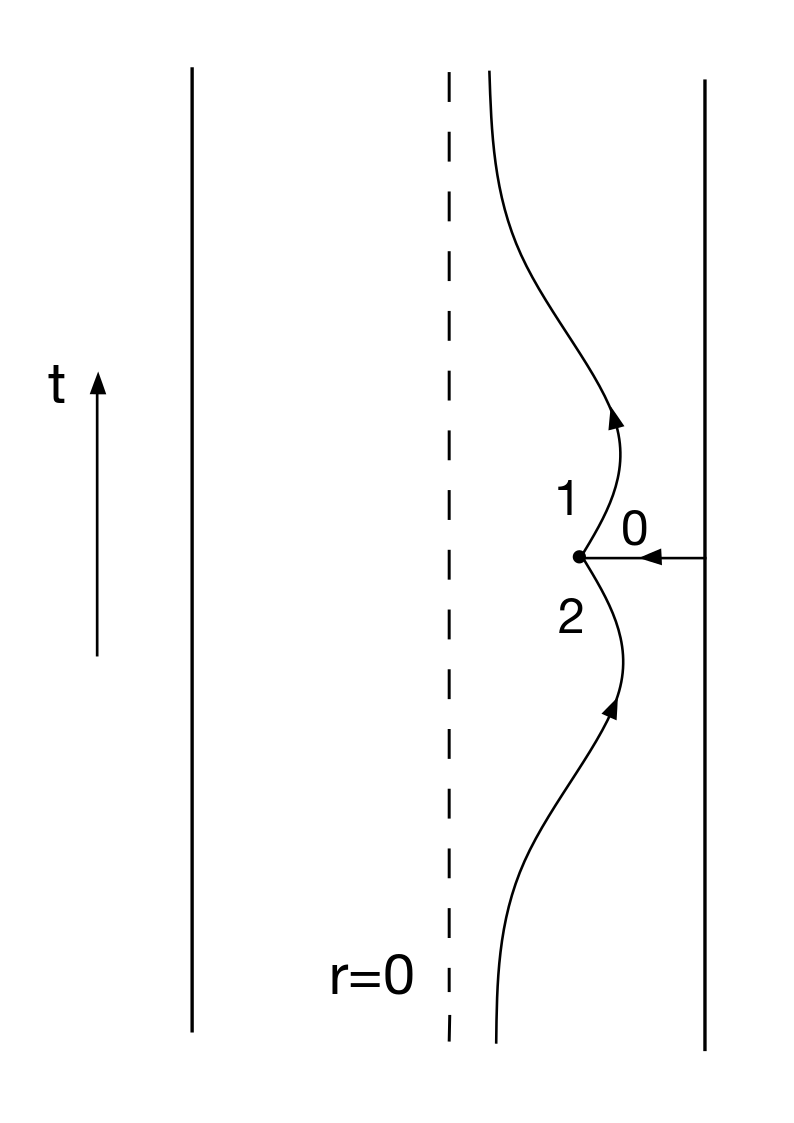} &   \includegraphics[width=40mm]{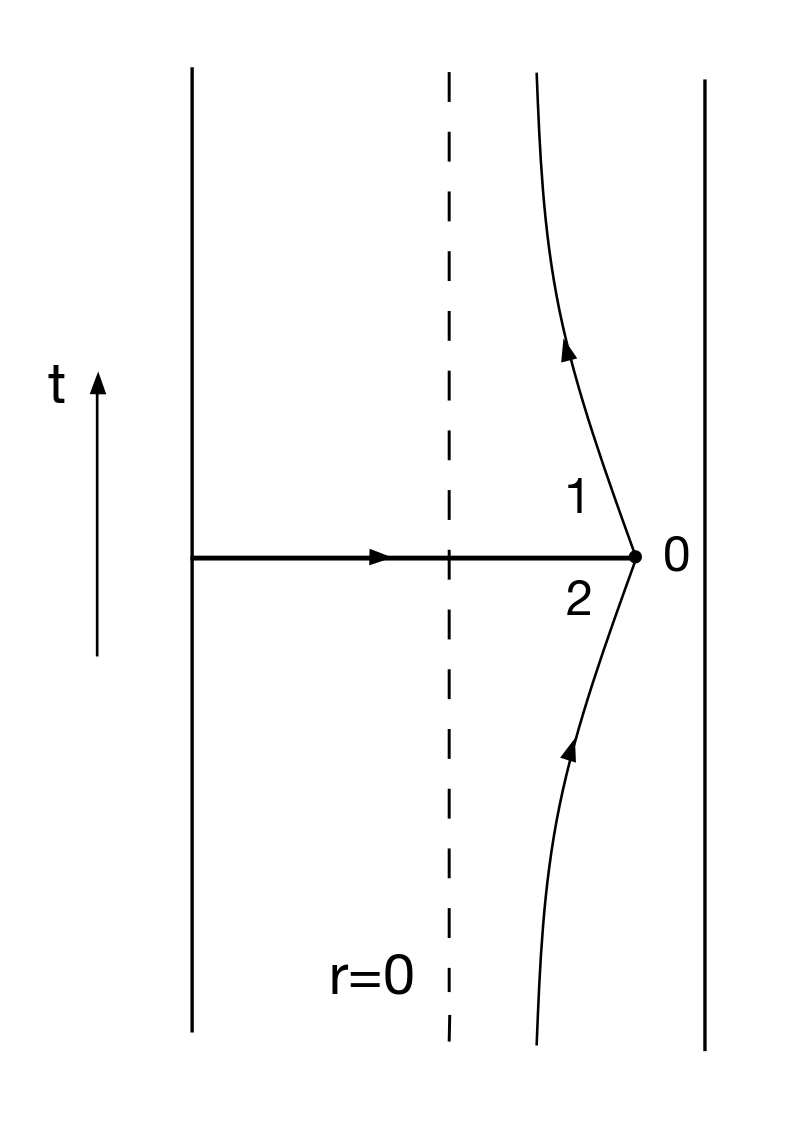} \\
(a)  & (b)  & (c) & (d)
\end{tabular}
\caption{Geodesics behaviour depending on the sign of the intermediate dimension. It is shown in the text that only the options $(a)$ and $(d)$ are realized, when $\epsilon_p>0$ and $\epsilon_p<0$ respectively. Arrows show the orientation of geodesics, chosen in the text.}
\label{fig:vertices}
\end{figure}
\begin{equation}
\label{eq:signs}
\begin{split}
& \operatorname{a)}\ \epsilon_p >0,\quad \dot r_0 <0,\quad \dot r_2>0\quad \quad \quad \quad \operatorname{c)}\ \epsilon_p <0, \quad \dot r_0 <0,\quad \dot r_2<0 \\ 
& \operatorname{b)}\ \epsilon_p >0, \quad \dot r_0 >0,\quad \dot r_2<0 \quad \quad \quad \quad \operatorname{d)}\ \epsilon_p <0, \quad \dot r_0 >0,\quad \dot r_2>0
\end{split}
\end{equation}
Soon it will be shown that options $\operatorname{(b)}$ and $\operatorname{(c)}$ can not be realized, as it follows from the geodesic equation that the inequality $\dot r_2>0$ holds for the loop (otherwise the geodesic corresponding to the intermediate dimension reaches infinity $r=\infty$ and does not form a loop). From the equation (\ref{eq:momentum}) one finds the radial coordinate of the vertex as the function of the momentum vector component $p_t$, conserved along the loop:
\begin{equation}
\label{eq:vertex}
\rho=\sqrt{\frac{s^2}{1-\delta^2/4}-1},
\end{equation}
where the following notations are introduced: $\delta=\epsilon/\epsilon_p,\, s=p_t/|\tau|$. As was mentioned above, the whole geodesic configuration exists only when $|\delta|<2$. From the equations (\ref{eq:geodes}) one obtains the law of radial motion: 
\begin{equation}
\label{eq:newton}
\left(\frac{d r}{dt}\right)^2+V(r)\equiv\left(\frac{d r}{dt}\right)^2-|\tau|^2\left(1+r^2\right)^2\,\frac{1+r^2-s^2}{s^2}=0.
\end{equation}
The qualitative behaviour of the effective potential $V(r)$ depends on whether $s^2>1$ or $s^2<1$. If $s^2<1$, then the particle reaches the radial coordinate $r=0$. In this case the smooth geodesic does not form a loop, so we do not consider it in what follows. On the other hand, if $s^2>1$, the particle moving in the effective potential $V(r)$ with zero energy changes direction of its motion at the point $r=r_{*}\equiv \sqrt{s^2-1}$. In this case we find the general solution to the equation (\ref{eq:newton}) $t(r)$, depending on two constants $t_0$ and $s$:
\begin{equation}
\label{eq:case2}
e^{2|\tau|(t-t_0)}=-\frac{-r\left[2s\sqrt{1+r^2-s^2}+r(1+s^2)\right]+s^2-1}{(r^2+1)(s^2-1)},    
\end{equation}
and to fulfill the condition $r(t=0)=r_*$ (otherwise the boundary condition for the loop (\ref{eq:bound_cond_loop}) can not be satisfied) one must set the constant $t_0$ to $0$. For the requirement $r(\pi)=\rho$ to be satisfied, the conserved momentum $s$ must be expressed through the radial coordinate of the vertex $\rho$ as follows:
\begin{equation}
\label{eq:boundcondcase2}
e^{2|\tau|\pi}=-\frac{-\rho\left[2s\sqrt{1+\rho^2-s^2}+\rho(1+s^2)\right]+s^2-1}{(\rho^2+1)(s^2-1)}.
\end{equation}
The solution (\ref{eq:case2}), (\ref{eq:boundcondcase2}) describes a particle, starting its motion from the radial coordinate $r=\rho$ at the moment  $t=-\pi$ and reaching the coordinate $r=r_{*}$ at the moment $t=0$. Let us also note that $\rho=\sqrt{\frac{s^2}{1-\delta^2/4}-1}\ge r_*=\sqrt{s^2-1}$, so the configuration consisting of the leg and the loop always exists. Substituting $\rho (s)$ from the vertex condition (\ref{eq:vertex}) into the equation (\ref{eq:boundcondcase2}), one obtains the following equation for the conserved momentum $s$, defining the trajectory of a particle, moving along the loop:
\begin{equation}
\label{eq:1var}
q^{-1}= e^{2|\tau|\pi}= 1+\frac{\delta^2/2+|\delta|\sqrt{s^2-1+\delta^2/4}}{s^2-1},
\end{equation}
which is solved as follows:
\begin{equation}
\label{eq:momentumofdelta}
s^2=1+\frac{\delta^2q^{-1}}{(q^{-1}-1)^2}.  
\end{equation}
Thus, the desired configuration of geodesics is found, and what is left is to calculate lengths of both geodesics:
\begin{equation}
S_{\operatorname{loop}}=2\int_{r_{*}}^{\rho} \frac{dr}{\sqrt{1+r^2-s^2}}=2\log \frac{\rho+\sqrt{1+\rho^2-s^2}}{\sqrt{s^2-1}},
\end{equation}
\begin{equation}
S_{\operatorname{leg}}=
\begin{cases}
\displaystyle \int_{\rho}^{\Lambda} \frac{dr}{\sqrt{1+r^2}} =-\arsinh\sqrt{\frac{s^2}{1-\delta^2/4}-1},\\[10pt]
\displaystyle \int_{0}^{\Lambda} \frac{dr}{\sqrt{1+r^2}}+\int_{0}^{\rho} \frac{dr}{\sqrt{1+r^2}} =\arsinh\sqrt{\frac{s^2}{1-\delta^2/4}-1},
\end{cases}
\end{equation}
where two options for the leg length correspond to the cases of positive and negative intermediate dimensions respectively, i.e. to configurations $(a)$ and $(d)$ in Fig. \ref{fig:vertices}. The regulator $\Lambda$ tends to $\infty$, and we discard those terms, which depend only on $\Lambda$, in the final answer. Taking into account the equation (\ref{eq:vertex}), defining the position of the vertex, and the equation (\ref{eq:momentumofdelta}), defining the momentum in the loop $s$ as the function of $\delta$ and $q$, we find:
\begin{equation}
\label{eq:lengths_final_torus}
\begin{split}
&S_{\operatorname{loop}}=2\log \frac{\left|q^{-1}+1\right|+\sqrt{(q^{-1}-1)^2+\delta^2 q^{-1}}  }{\sqrt{q^{-1}}\sqrt{4-\delta^2}},\\
&S_{\operatorname{leg}}=\mp\log \frac{|\delta|\left|q^{-1}+1\right|+2\sqrt{(q^{-1}-1)^2+\delta^2 q^{-1}}  }{\left|q^{-1}-1\right|\sqrt{4-\delta^2}},
\end{split}
\end{equation}
So, we obtain the following closed-form expression for the linearized semiclassical conformal block on the torus\footnote{Despite the fact that the holographic picture only exists for $|\delta|<2$, the linearized semiclassical conformal block on the torus exists for all values of $\delta$. When $\delta \to \pm 2$, the vertex radial coordinate (\ref{eq:vertex}) tends to infinity, and when $|\delta|>2$, it becomes imaginary, so that the holographic picture breaks down. However, as a function of $\delta$, the holographic action is equal to the conformal block for all values of $\delta\in \mathbb{R}$, which can be checked by expanding the former in powers of $q$.}
for $q<1$, taking into account that $f_{\, \operatorname{tor}}^{\, \operatorname{lin}}(\epsilon_p,\epsilon|0)=0$:
\begin{equation}
\label{eq:torus_answer}
f_{\, \operatorname{tor}}^{\, \operatorname{lin}}(\epsilon_p,\epsilon|q)= 2 \epsilon_p \log \frac{1+q+\sqrt{(1-q)^2+\delta^2 q }}{2}\mp\epsilon \log \frac{|\delta|\left(1+q\right)+2\sqrt{(1-q)^2+\delta^2 q}  }{\left(1-q\right)\left(|\delta|+2\right)},
\end{equation}
where the first sign should be taken for positive $\epsilon_p$, and vice versa. This expression has different limits when $\epsilon_p\to \pm 0$ (or, equivalently, when $\nu\to 1\mp 0$):
\begin{equation}
f_{\, \operatorname{tor}}^{\, \operatorname{lin}}(\epsilon_p,\epsilon|q)\to \mp \epsilon \log \frac{1+\sqrt{q}}{1-\sqrt{q}}.
\end{equation}
In conclusion, it is worth noting once again that geodesic configurations, reproducing the linearized semiclassical torus conformal block, qualitatively differ in the leg position in the cases $\epsilon_p>0$ and $\epsilon_p<0$ (Fig. \ref{fig:vertices}, $(a)$ and $(d)$). It is also useful to understand how conclusions about the analytic structure of the semicalssical irregular conformal block, made in section \ref{irreg} on the basis of its double series representation, and the analytic structure of the linearized torus conformal block (\ref{eq:torus_answer}) are related to each other. First, in the case of the irregular conformal block we dealt with a function having the expansion in $q$ of the form (\ref{eq:irreg_block_exp}), valid for both $\nu>1$ and $\nu<1$, and having different limits from the right and from the left when $\nu\to 1$. The same is true for the torus conformal block, because the formula (\ref{eq:torus_answer}) can be uniformly written for both positive and negative intermediate dimensions $\epsilon_p$:
\begin{equation}
\label{eq:torus_one}
f_{\, \operatorname{tor}}^{\, \operatorname{lin}}(\epsilon_p,\epsilon|q)= 2 \epsilon_p \log \frac{1+q+\sqrt{(1-q)^2+\delta^2 q }}{2}-\epsilon \log \frac{\delta\left(1+q\right)+2\sqrt{(1-q)^2+\delta^2 q}  }{\left(1-q\right)\left(\delta+2\right)}.
\end{equation}
Second, square roots and logarithms, appearing after the resummation of the irregular conformal block (\ref{eq:resum}), are exactly the same as square roots and logarithms, appearing after the resummation of the torus conformal block (\ref{eq:torus_resum}). Comparing the resummed series (\ref{eq:torus_resum}) and the analytic answer (\ref{eq:torus_one}) for the linearized torus block, we see that the resummed series captures a significant part of the analytic structure of the conformal block. Of course, the expansion (\ref{eq:torus_resum}) can now be easily obtained from the analytic answer (\ref{eq:torus_one}).

\subsection{Generalization of holographic picture to other poles}
It is worth mentioning that there also exist tadpole configurations, solving the equations of motion and satisfying the momentum conservation law in the vertex, for which the loop wraps the torus $n$ times, $n\in \mathbb{N}$. In this case the boundary condition for the loop changes (\ref{eq:bound_cond_loop}):
\begin{equation}
\label{eq:bound_cond_loop_gen}
r(-n\pi)=r(n\pi)=\rho,    
\end{equation}
which leads to the change of the equation (\ref{eq:1var}):
\begin{equation}
q^{-n}\equiv e^{2|\tau|n\pi}= 1+\frac{\delta^2/2+|\delta|\sqrt{s^2-1+\delta^2/4}}{s^2-1}.    
\end{equation}
Thus, in the case of $n$ windings one has to make the replacement $q\to q^n$ in the formula (\ref{eq:lengths_final_torus}) for the loop and leg lengths. A natural question arises: what corresponds to the geodesic configuration with the winding number $n$ in CFT${}_2$? The action of this configuration, where the mass of the particle in the loop equals  $\epsilon_p-\epsilon_n\equiv \epsilon_p-(1-n^2)/4$ and the mass of the particle in the leg equals $\epsilon$, reproduces the $n$th linearized torus conformal block $f_{\ \operatorname{tor}}^{(n)}(\epsilon_p,\epsilon|q)$, which is obtained from the full semiclassical 1-point conformal block on the torus in the following limit (compare to the limit (\ref{eq:linear_toric})):   
\begin{equation}
\epsilon_p - \epsilon_n\to 0,\ \epsilon\to 0,\ \text{while}\ \delta_n\equiv \frac{n\epsilon}{\epsilon_p-\epsilon_n}\to \operatorname{const}.  
\end{equation}
From the perturbative expansion of the conformal block on the torus (\ref{eq:toricblock_coeff}), we find the following non-zero coefficients of the expansion of the $n$th linearized conformal block in $q$ (compare to (\ref{eq:linearblock_coeff})):
\begin{equation}
\label{eq:linearblock_coeff_nth}
\begin{split}
&f^{(n)}_{\,n}(\epsilon_p,\epsilon)=\frac{\epsilon_p-\epsilon_n}{n}\left( -\frac{\delta_n^2}{2}\right)=-\frac{\epsilon^2}{n-\nu},\\
&f^{(n)}_{\,2n}(\epsilon_p,\epsilon)=\frac{\epsilon_p-\epsilon_n}{n}\left(\frac{-8\delta_n^2+\delta_n^4}{16}\right)=\frac{\epsilon^4}{2 (n - \nu)^3} - \frac{
\epsilon^2}{(n - \nu)},\\
&f^{(n)}_{\,3n}(\epsilon_p,\epsilon)=\frac{\epsilon_p-\epsilon_n}{n}\left( \frac{-24 \delta_n^2 + 8 \delta_n^4 - \delta_n^6}{48}\right)=-\frac{2 \epsilon^6}{3 (n-\nu)^5} + \frac{4 \epsilon^4}{3 (n-\nu)^3}- \frac{\epsilon^2}{n - \nu},
\end{split}
\end{equation}
where the equality $\epsilon_p-\epsilon_n=n(n-\nu)/2$ has been used, which follows from (\ref{eq:dim_torus}) in the limit $\nu\to n$. Based on the first expansion coefficients of the $n$th linearized conformal block, one concludes that
\begin{equation}
f_{\, \operatorname{tor}}^{(n)}(\epsilon_p,\epsilon|q)=f_{\, \operatorname{tor}}^{\, \operatorname{lin}}\left(\frac{\epsilon_p-\epsilon_n}{n},\epsilon\bigg|q^n\right),   
\end{equation}
where the r.h.s of this equation coincides with the length of the geodesic configuration with the winding number $n$, mentioned above. 
 
\section{Four-point conformal block on sphere}\label{4pt}
\subsection{Heavy-light limit of semiclassical four-point conformal block on sphere}
In this section we briefly discuss the so called heavy-light limit of the semiclassical 4-point conformal block on the sphere. By definition, the semiclassical heavy-light conformal block is obtained from the semiclassical 4-point conformal block on the sphere 
\begin{equation}
f_{\operatorname{4pt}}(\epsilon_p,\epsilon_i|z)=\lim_{b\to 0} b^2 \log \mathcal F_{\operatorname{4pt}}(\Delta_p,\Delta_i,c|z)    
\end{equation}
in the limit $\epsilon_p\to 0,\, \epsilon_{1,2}\to 0$, where all posible ratios of these semiclassical conformal dimensions are held fixed \cite{HijanoKraus}. In the equality above the following notations for conformal dimensions in the semiclassical limit, analogous to (\ref{eq:clas_parametr_mathieu}) and (\ref{eq:dim_torus}), were introduced:
\begin{equation}
\label{eq:clas_parametr_heun}
\epsilon_i\equiv b^2\Delta_i\big|_{b\to 0},\ \epsilon_p\equiv    b^2\Delta_p\big|_{b\to 0}.    
\end{equation}
More precisely, the semiclassical 4-point heavy-light conformal block is the first (linear) order in the expansion of the ordinary semiclassical 4-point conformal block in parameters $\epsilon_p,\, \epsilon_1$ and $\epsilon_2$, with the condition on their ratios mentioned above being fulfilled. So, the heavy-light limit of the semiclassical conformal block on the sphere is an analogue of the linearized limit of the semiclassical torus conformal block (\ref{eq:linear_toric}), considered before. Let us note that all the conformal dimensions entering the heavy-light conformal block are actually heavy ($\Delta_i,\, \Delta_p \sim b^{-2}$ when $b\to 0$), despite of its name. The reason, why some of them are called "heavy"\ and others "light"${}$, lies in the fact that $\epsilon_p,\, \epsilon_{1,2} \ll 1$. In what follows the case of equal "heavy"\ dimensions is considered $\epsilon_3=\epsilon_4=\epsilon_h$, and the following notation is introduced for "light"\ dimensions: $\epsilon_1=\epsilon_l,\, \epsilon_2=\epsilon_l'$.   

\subsection{Holographic interpretation of semiclassical conformal block on the sphere }
The holographic interpretation of the non-vacuum semiclassical 4-point heavy-light conformal block on the sphere in terms of geodesics in the conical defect background and in the BTZ black hole background was suggested in the work \cite{HijanoKraus}. Namely, it is equal to the extremum of the action of the configuration of three geodesics in the background of the conical defect in AdS${}_3$ (Fig. \ref{fig:defect}) or of the BTZ black hole (Fig. \ref{fig:btz2}):
\begin{figure}
\centering
\includegraphics[scale=0.25]{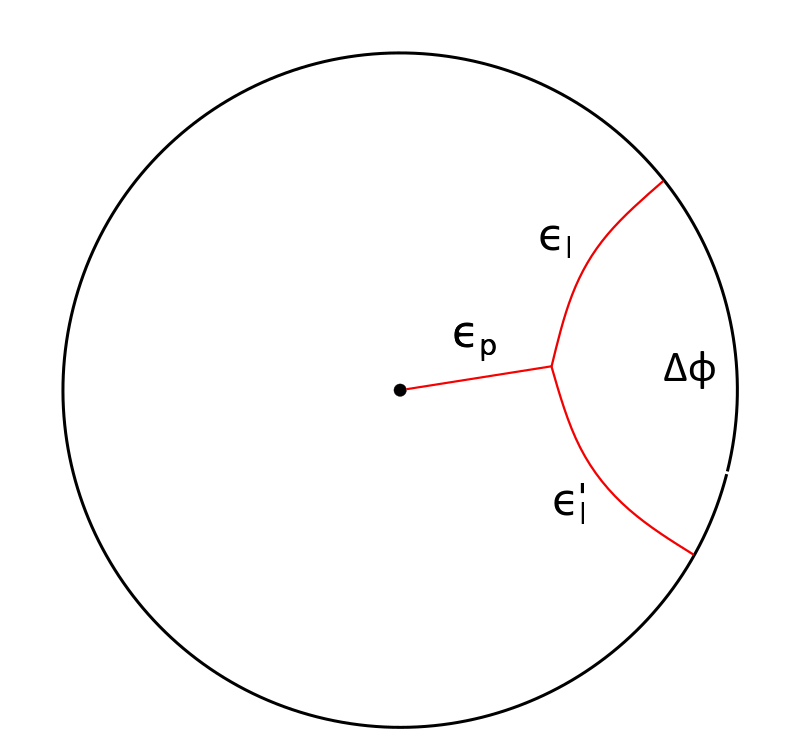}
\caption{The configuration of geodesics in the background of the conical defect in AdS${}_3$, which is dual to the semiclassical 4-point heavy-light conformal block on the sphere. The disc depicted is a slice of AdS${}_3$ at constant time $t=\operatorname{const}$. Polar coordinates $(\rho, \phi)$ are introduced on the disc.}
\label{fig:defect}
\end{figure}
\begin{equation}
    \label{eq:total_act_4pt}
f(\epsilon_p,\epsilon_i|z)=\epsilon_l \log(1-z)+S^{\operatorname{(extr)}}(\epsilon_p,\epsilon_i|z)=\epsilon_l\log(1-z)+\left(\epsilon_pS_p+\epsilon_l S_l+\epsilon_l' S_l'\right)^{\operatorname{(extr)}},
\end{equation}
where $z$ is related to the position of endpoints of the geodesics attached to the boundary  (the precise relation is given below). "Heavy"\ external dimensions, which are equal to each other, define the background geometry through the parameter $\alpha=\sqrt{1-4\epsilon_h}$. When $\epsilon_h<1/4$ ($\alpha$ is real), one gets the conical defect, and when $\epsilon_h>1/4$ ($\alpha$ is imaginary), one gets the BTZ black hole. "Light"\ external dimensions define masses of particles that move along geodesics attached to the boundary of the space at one end. The intermediate dimension is equal to the mass of the particle, whose trajectory ends at the singularity or at the horizon. 
%either the conical defect or a BTZ black hole $\alpha=\sqrt{1-4\epsilon_h}\ $ in AdS${}_3$.

We generalize the holographic correspondence for geodesics in the conical defect background to the case of negative intermediate dimensions and also find the geodesic configuration at a constant time slice in the BTZ black hole background, which is dual to the conformal block for imaginary $\alpha$. We show that, depending on the sign of the intermediate dimension $\epsilon_p$, the extremum geodesic configuration is defined by different branches of a square root. Using the analytic answer, we find the same discontinuity in this conformal block as we had above for the other conformal blocks. 

In appendix \ref{AppGlobalAdS} we also show that the configurations of geodesics in the BTZ black hole and in the conical defect backgrounds, mentioned above, map into the same configuration when one embeds the BTZ black hole and the conical defect into global AdS$_3$ space. In appendix \ref{AppBTZ_const_angle} we reproduce the heavy-light conformal block considering geodesics at a constant angle in the BTZ black hole background and explicitly show that they are obtained by analytic continuation from geodesics at constant time in the conical defect background. However, it does not seem possible to find a geodesic configuration in this background corresponding to $\epsilon_p<0$ and thus to reproduce the discontinuity this way.

\subsubsection{Conical defect}\label{Con_def}
%% In this section we generalize their result to the case of negative intermediate dimensions and find the same discontinuity in this conformal block as we had above for the other conformal blocks. 
%The semiclassical 4-point heavy-light conformal block on the sphere is equal to the extremum of the action for the configuration consisting of three geodesics in the background of the conical defect in AdS${}_3$ (Fig. \ref{fig:defect}):
%begin{equation}
%\label{eq:total_act_4pt}
%S=\epsilon_pS_p+\epsilon_l S_l+\epsilon_l' S_l'.
%\end{equation}

%"Light"\ external dimensions define masses of particles moving along geodesics attached at one end to the boundary of AdS${}_3$. The angle between end points of these geodesics $\Delta \phi$ is identified with the coordinate $z$ in the conformal block (see below). The intermediate dimension is equal to the mass of the particle, the trajectory of which ends at the singularity $\rho=0$. "Heavy"\ external dimensions, which are equal to each other, define the conical defect $\alpha=\sqrt{1-4\epsilon_h}\ $ in AdS${}_3$,
In this section we demonstrate how the semiclassical heavy-light 4-point conformal block on the sphere and its discontinuity are reproduced holographically in terms of geodesics at constant time in the conical defect background. 

The metric of the conical defect has the following form:
\begin{figure}
\begin{tabular}{cc}
  \includegraphics[width=84mm]{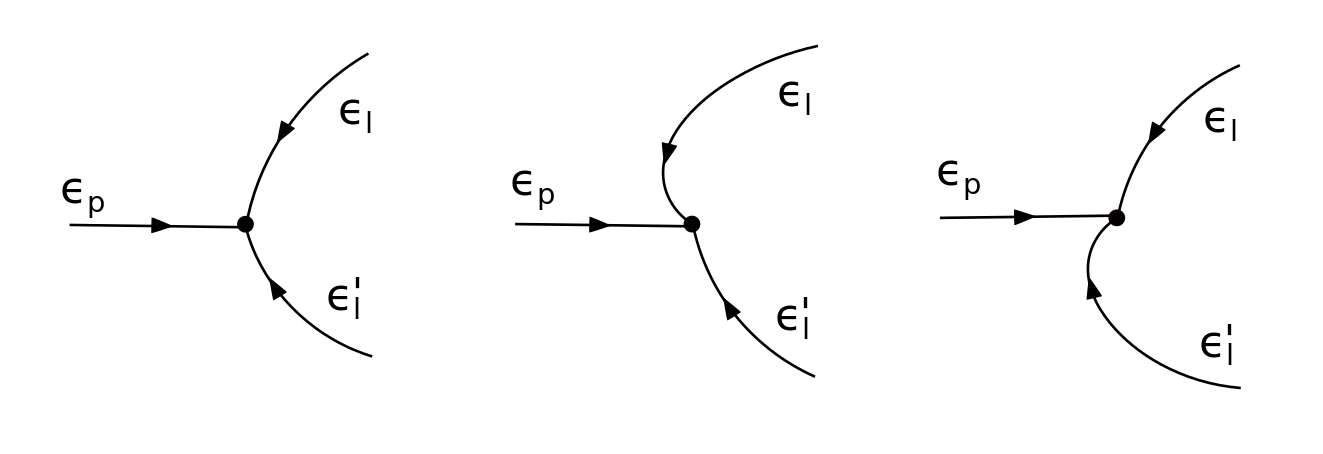} &   \includegraphics[width=84mm]{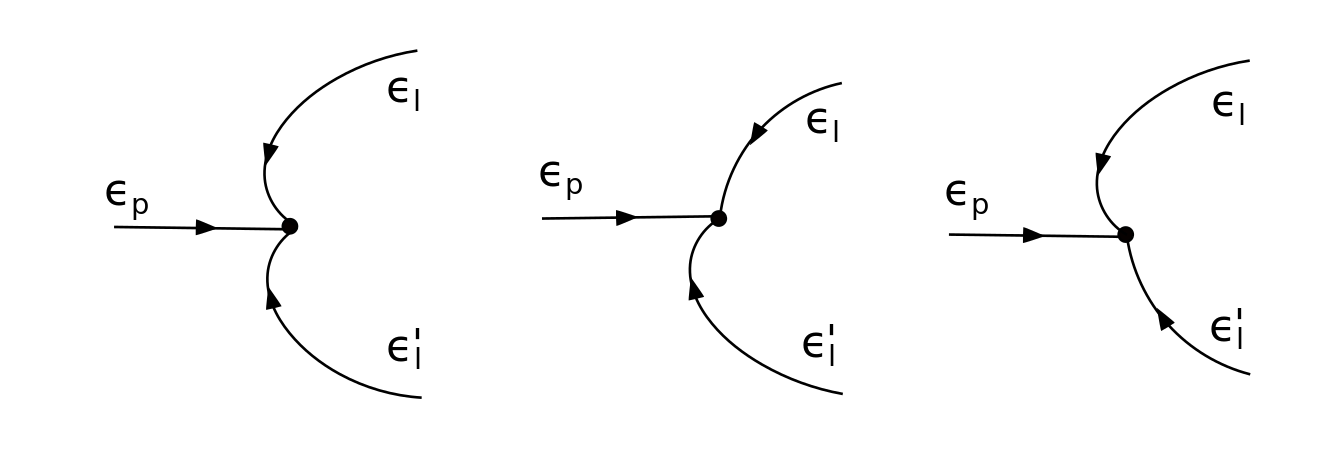} \\
(a) $\epsilon_p>0$, configurations $--,+-,-+$ & (b) $\epsilon_p<0$, configurations $--,-+,+-$
\end{tabular}
\caption{Possible behaviour of geodesics in the vicinity of the vertex. Arrows show the orientation of geodesics, chosen in the text.}
\label{fig:vertices_4pt}
\end{figure}
\begin{equation}
\label{eq:metric_defect}
ds^2=\frac{\alpha^2}{\cos^2 \rho}\left(\frac{1}{\alpha^2}d\rho^2-dt^2+\sin^2\rho\, d\phi^2\right),
\end{equation}
where $0\le\rho<\pi/2$, $-\infty<t<+\infty$, and $\phi\sim \phi+2\pi$. Only those geodesics are considered, which belong to the slice $t=\operatorname{const}$ of the original cylinder (Fig. \ref{fig:defect}). Choosing the parametrization of geodesics in such a way that for each of them velocity vectors are everywhere directed toward the vertex, one obtains the law of conservation of momentum in the vertex in the following form:
\begin{equation}
\label{eq:mom_cons}
\begin{split}
&\epsilon_lp_\phi+\epsilon_l'p_\phi'=0,\\
&\pm \epsilon_l \sqrt{1-\frac{p_\phi^2}{\alpha^2 \tan^2\rho_v}}\pm \epsilon_l' \sqrt{1-\frac{p_\phi'^2}{\alpha^2 \tan^2\rho_v}}+\epsilon_p=0,
\end{split}
\end{equation}
where $p_\phi=g_{\phi\phi}\,d\phi/ds=\operatorname{const}$ is the angular momentum, conserved along the trajectory of the particle with the mass $\epsilon_l$ (the definition of $p_\phi'$ is analogous). For a positive intermediate dimension $\epsilon_p>0$ (we consider only positive "light"\ external dimensions, such that $\epsilon_l\ge \epsilon_l'>0$), there are 3 options to choose the sign in front of $\epsilon_l$ and $\epsilon_l'$ respectively: $--,\ +-,\ -+$. When the intermediate dimension is negative $\epsilon_p<0$, there are the following options: $++,\ -+,\ +-$. In both cases the first option is realized when $\epsilon_l^2-\epsilon_l'^2\le \epsilon_p^2$, the second one is impossible, and the third one is realized when $\epsilon_l^2-\epsilon_l'^2\ge \epsilon_p^2$. Acoording to the chosen parametrization of geodesics, the plus sign in front of the square root means that the geodesic approaches the vertex from inward, and vice versa (Fig. \ref{fig:vertices_4pt}). From the momentum conservation law (\ref{eq:mom_cons}), we find the radial coordinate of the vertex $\rho_v$ as a function of the angular momentum $p_\phi$, conserved along the leg:
\begin{equation}
\label{eq:vertex_def}
\frac{(\epsilon_lp_\phi)^2}{\alpha^2 \tan^2 \rho_v}= \mu^2\equiv \frac{\epsilon_l^2+\epsilon_l'^2-\epsilon_p^2/2}{2}-\frac{(\epsilon_l^2-\epsilon_l'^2)^2}{4\epsilon_p^2}.    
\end{equation}
Note that $\mu^2\ge0$ if and only if $\epsilon_l-\epsilon_l'\le |\epsilon_p|\le\epsilon_l+\epsilon_l'$, so the holographic picture exists only when this inequality is satisfied. From the equation $u^2=1$ and angular momentum conservation one obtains the following law of radial motion:
\begin{equation}
\label{eq:defect_eff_potential}
\left(\frac{d\rho}{d\phi}\right)^2+V(r)\equiv\left(\frac{d\rho}{d\phi}\right)^2-\alpha^2\sin^2 \rho \left(\frac{\alpha^2 \tan^2 \rho}{p_\phi^2}-1\right)=0. 
\end{equation}
For a particle with zero energy the effective potential $V(r)$ has a reflection point $\rho_*=\arctan |p_\phi|/\alpha$. From the last equation one can find the angle between end points of external legs $\Delta \phi$ as a function of the conserved angular momentum $p_\phi$ for configurations $--$ and $-+$ in the case of positive intermediate dimensions $\epsilon_p>0$:
\begin{equation}
\label{eq:dphi12}
\begin{split}
&\Delta \phi^{--}=\frac{ |p_\phi|}{\alpha^2}\int_{\rho_v}^{\pi/2}\frac{d\rho}{\sin \rho \sqrt{ \tan^2 \rho-(p_\phi/\alpha)^2}}+\frac{ |p_\phi'|}{\alpha^2}\int_{\rho_v}^{\pi/2}\frac{d\rho}{\sin \rho \sqrt{\tan^2 \rho-(p_\phi'/\alpha)^2}}=\frac{1}{\alpha}\bigg[\arcsin \frac{\sin\rho_*}{\sin \rho_v}-\\
&-\rho_*+\arcsin \frac{\sin\rho_*'}{\sin \rho_v}-\rho_*'\bigg],\\
&\Delta \phi^{-+}=\Delta \phi^{--}+\frac{2 |p_\phi'|}{\alpha^2}\int_{\rho_*'}^{\rho_v}\frac{d\rho}{\sin \rho \sqrt{ \tan^2 \rho-(p_\phi'/\alpha)^2}}=\frac{1}{\alpha}\left[\arcsin \frac{\sin\rho_*}{\sin \rho_v}-\rho_*+\pi-\arcsin \frac{\sin\rho_*'}{\sin \rho_v}-\rho_*'\right],
\end{split}
\end{equation}
where $\rho_*$ and $\rho_*'$ are reflection points for particles with masses $\epsilon_l$ and $\epsilon_l'$ respectively. Analogously, one finds the angle $\Delta \phi$ for configurations $++$ and $+-$ in the case of negative intermediate dimensions $\epsilon_p<0$:
\begin{equation}
\label{eq:dphi34}
\begin{split}
&\Delta \phi^{++}=\frac{1}{\alpha}\left[2\pi-\arcsin \frac{\sin\rho_*}{\sin \rho_v}-\rho_*-\arcsin \frac{\sin\rho_*'}{\sin \rho_v}-\rho_*'\right],\\
&\Delta \phi^{+-}=\frac{1}{\alpha}\left[\pi-\arcsin \frac{\sin\rho_*}{\sin \rho_v}-\rho_*+\arcsin \frac{\sin\rho_*'}{\sin \rho_v}-\rho_*'\right].
\end{split}
\end{equation}
From the equations (\ref{eq:dphi12}), (\ref{eq:dphi34}) one finds the radial coordinate of the vertex $\rho_v$ as a function of the angle $\theta\equiv \alpha\Delta\phi/2$ for all the configurations and both signs of the intermediate dimension:
\begin{equation}
\label{eq:vertex_pos}
\tan \rho_v=\frac{\epsilon_l+\epsilon_l'}{2\mu}\cot\theta\mp\frac{\sqrt{\epsilon_p^2-(\epsilon_l-\epsilon_l')^2\sin^2\theta}}{2\mu\sin \theta}, 
\end{equation}
where the first sign corresponds to the positive intermediate dimension, and vice versa (both of the equations (\ref{eq:dphi12}) have the same solution as well as both of the equations (\ref{eq:dphi34})). Instead of calculating geodesic lengths directly, it is much simpler to find how the total action (\ref{eq:total_act_4pt}) changes with the change of the angle $\theta$ \cite{HijanoKraus}:
\begin{equation}
\label{eq:reg_length_def}
\frac{dS}{d\theta}=2\mu \tan \rho_v=(\epsilon_l+\epsilon_l')\cot\theta \mp \frac{\sqrt{\epsilon_p^2-(\epsilon_l-\epsilon_l')^2\sin^2\theta}}{\sin \theta}.
\end{equation}
Integrating this equation, we find the action of the configuration of three geodesics: 
\begin{equation}
\label{eq:action_defect}
S=(\epsilon_l+\epsilon_l')\log\sin \theta+\epsilon_p\artanh\frac{\cos \theta}{\sqrt{1-\beta^2\sin^2\theta}}\mp(\epsilon_l-\epsilon_l')\log\left(|\beta|\cos\theta+\sqrt{1-\beta^2\sin^2\theta}\right),
\end{equation}
where $\beta\equiv(\epsilon_l-\epsilon_l')/\epsilon_p$ and the angle $\theta$ is identified with the coordinate $z$ in the conformal block as follows:
\begin{equation}
\begin{split}
&\sin \theta=-i\sinh \frac{\alpha\log(1-z)}{2}\equiv-i\sinh \theta', \\   
&\cos \theta=\cosh \frac{\alpha\log(1-z)}{2}\equiv\cosh \theta'. 
\end{split}    
\end{equation}
Thus, we obtain the following closed-form expression for the heavy-light conformal block for both signs of the intermediate dimension\footnote{Despite the fact that the holographic picture exists only when $\epsilon_l-\epsilon_l'\le |\epsilon_p|\le\epsilon_l+\epsilon_l'$, the heavy-light conformal block exists also for conformal dimensions not satisfying this inequality. So, as a function of $\epsilon_p,\,\epsilon_l,$ and $\epsilon_l'$, the holographic action is equal to the conformal block for all $\epsilon_p,\, \epsilon_l,\,\epsilon_l'\in \mathbb{R}$, which can be checked by expanding the former in powers of $z$. Exactly the same comment is also true for the next section, where the conformal block is reproduced as the action of geodesics at a constant time slice in the BTZ black hole background.}, taking into account that $f^{\, \operatorname{lin}}_{\, \operatorname{4pt}}(\epsilon_p,\epsilon_l,\epsilon_l'|0)=0$:
\begin{equation}
\label{eq:4pt_answer}
\begin{split}
&f^{\, \operatorname{lin}}_{\, \operatorname{4pt}}(\epsilon_p,\epsilon_l,\epsilon_l'|z)=\epsilon_l \log(1-z)+(\epsilon_l+\epsilon_l')\log\left(-\sinh \theta'\right)-(\epsilon_l+\epsilon_l')\log \frac{\alpha z}{2}+\\
&+\frac{\epsilon_p}{2}\log\frac{\cosh\theta'+\sqrt{1+\beta^2\sinh^2\theta'}}{\cosh\theta'-\sqrt{1+\beta^2\sinh^2\theta'}}+\epsilon_p\log \frac{\sqrt{1-\beta^2}\,\alpha z}{4}\mp (\epsilon_l-\epsilon_l')\log\frac{|\beta|\cosh\theta'+\sqrt{1+\beta^2\sinh^2\theta'}}{|\beta|+1}, 
\end{split}
\end{equation}
where the first sign should be taken for positive $\epsilon_p$, and vice versa. This expression has different limits when $\epsilon_p\to \pm 0$:
\begin{equation}
f^{\, \operatorname{lin}}_{\, \operatorname{4pt}}(\epsilon_p,\epsilon_l,\epsilon_l'|z)\big|_{\epsilon_p\to +0}-f^{\, \operatorname{lin}}_{\, \operatorname{4pt}}(\epsilon_p,\epsilon_l,\epsilon_l'|z)\big|_{\epsilon_p\to -0}=2 (\epsilon_l'-\epsilon_l)\,|\theta'|.
\end{equation}
As is seen from the last formula, the difference between values of the heavy-light conformal block at points $\nu=1+0$ and $\nu=1-0$ is proportional to the difference between "light"\ external dimensions, so when $\epsilon_l=\epsilon_l'$, one obtains from (\ref{eq:4pt_answer}) an unambiguous linearized vacuum block, setting $\nu=1$. As in the case of the linearized torus block, the answer for the heavy-light conformal block can be uniformly written for both positive and negative intermediate dimensions:
\begin{equation}
\begin{split}
&f^{\, \operatorname{lin}}_{\, \operatorname{4pt}}(\epsilon_p,\epsilon_l,\epsilon_l'|z)=\epsilon_l \log(1-z)+(\epsilon_l+\epsilon_l')\log\left(-\sinh \theta'\right)-(\epsilon_l+\epsilon_l')\log \frac{\alpha z}{2}+\\
&+\frac{\epsilon_p}{2}\log\frac{\cosh\theta'+\sqrt{1+\beta^2\sinh^2\theta'}}{\cosh\theta'-\sqrt{1+\beta^2\sinh^2\theta'}}+\epsilon_p\log \frac{\sqrt{1-\beta^2}\,\alpha z}{4}- (\epsilon_l-\epsilon_l')\log\frac{\beta\cosh\theta'+\sqrt{1+\beta^2\sinh^2\theta'}}{\beta+1}. 
\end{split}
\end{equation}

\subsubsection{BTZ black hole. Geodesics at constant time}\label{BTZ_const_time}
In this section we demonstrate how the semiclassical heavy-light 4-point conformal block on the sphere and its discontinuity are reproduced holographically in terms of geodesics at constant time in the BTZ black hole background. 
\par When $\epsilon_h>1/4$, the parameter $\alpha\equiv\sqrt{1-4\epsilon_h}$, defining the background geometry (\ref{eq:metric_defect}), becomes imaginary and one obtains the metric of a non-rotating BTZ black hole outside the event horizon, substituting $\alpha=i\gamma$ into (\ref{eq:metric_defect}):   
\begin{equation}
\label{eq:BTZ_metric_1}
ds^2=\frac{\gamma^2}{\cos^2 \rho}\left(\frac{1}{\gamma^2}d\rho^2+dt^2-\sin^2\rho\, d\phi^2\right),
\end{equation}
where the range of coordinates is chosen as follows: $0\le\rho<\pi/2$, $-\infty<\phi<+\infty$, and $t\sim t+2\pi$. Indeed, after the transition to the Schwarzschild-like radial coordinate $r=\gamma/\cos\rho$, one obtains the standard BTZ black hole metric with swapped names for angular and time coordinates:
\begin{equation}
\label{eq:BTZ_metric_2}
ds^2=-(r^2-\gamma^2)d\phi^2+\frac{dr^2}{r^2-\gamma^2}+r^2dt^2,    
\end{equation}
so that the paramater $\gamma$ defines the event horizon position $r_+=\gamma$. In this section we rename the coordinates, so that $t$ is the time coordinate and $\phi$ is the angle:
\begin{equation}
ds^2=-(r^2-\gamma^2)dt^2+\frac{dr^2}{r^2-\gamma^2}+r^2d\phi^2,      
\end{equation}
where $0<r<\infty$, $-\infty<t<+\infty$, and $\phi\sim \phi+2\pi$. In the coordinates $(t,\rho,\phi)$ the event horizon is defined by the equation $\rho=0$, so we expect that the leg corresponding to the intermediate dimension should be attached to the horizon in order for the action of a geodesic configuration to reproduce the conformal block. We consider the geodesic configuration  on the constant time slice $t=\operatorname{const}$ (Fig. \ref{fig:btz2}) with the following induced metric:
\begin{figure}
\centering
\includegraphics[scale=0.25]{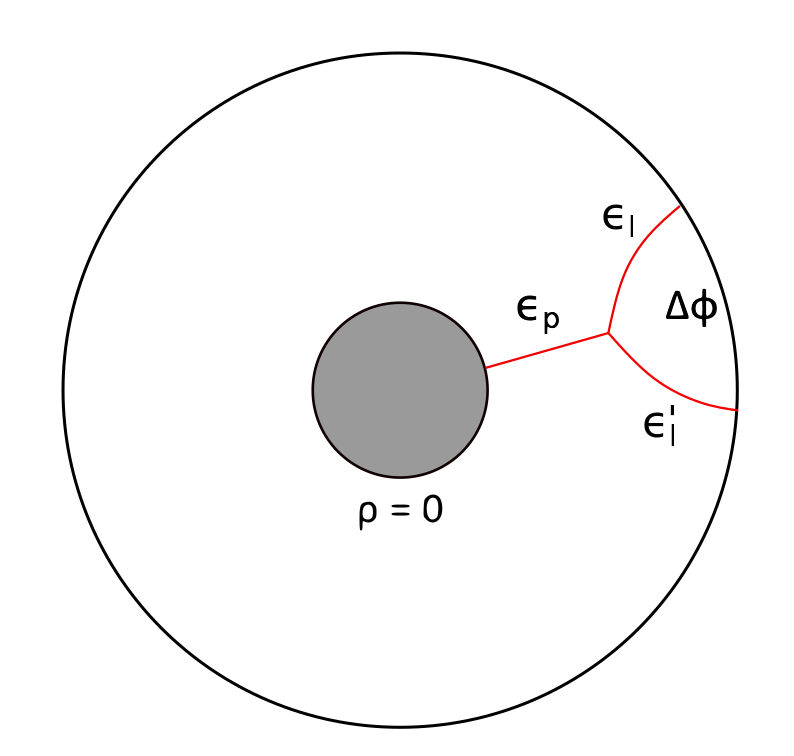}
\caption{The configuration of geodesics in the background of the BTZ black hole at the constant time slice, which is dual to the semiclassical 4-point heavy-light conformal block on the sphere. The smaller circle is the event horizon.}
\label{fig:btz2}
\end{figure}
\begin{equation}
    ds^2=\frac{1}{\cos^2\rho}\left(d\rho^2+\gamma^2d\phi^2\right).
\end{equation}
As above, from the equation $u^2=1$ and from the momentum conservation law $p_\phi\equiv g_{\phi\phi} d\phi/ds=\operatorname{const}$, one obtains the law of radial motion:
\begin{equation}
    \left(\frac{d\rho}{d \phi}\right)^2+\gamma^2\left(1-\frac{\gamma^2}{p_\phi^2\cos^2\rho}\right)=0
\end{equation}
If $p_\phi>\gamma$, then the effective potential has a turning point $\rho_*=\arccos \gamma/p_\phi$, so that the particle, launched from the boundary, returns there again after some time. Otherwise, it crosses the black hole event horizon. The general solution to the equation above, satisfying the condition $\rho(\phi_0)=\pi/2$, has the following form:
\begin{equation}
\label{eq:BTZ_motion_law}
    \phi-\phi_0=\frac{1}{\gamma}\log\frac{1+\gamma/p_\phi}{\sin \rho +\sqrt{\gamma^2/p_\phi^2-\cos^2\rho}}.
\end{equation}
The geodesic with the mass $\epsilon_p$ should be attached to the black hole horizon, but the question is where exactly at the horizon it should be attached. One might try to answer this question by embedding the BTZ black hole geometry to the global AdS${}_3$, where it is known that the end of the geodesic corresponding to the intermediate dimension must be placed at the "center" of the cylinder for the conformal block to be reproduced. Namely, one could try to attach this geodesic to some arbitrary point at the horizon. Instead, we consider geodesic configurations, where the geodesic with the mass $\epsilon_p$ satisfies the equation $d\phi=0$ (Fig. \ref{fig:btz2}). It solves the problem with the end point, described above, and, indeed, we will see that the conformal block is reproduced on such configurations. How these configurations relate to those considered in the previous subsection is discussed in appendix \ref{AppGlobalAdS}. The momentum conservation law in the vertex is analogous to (\ref{eq:mom_cons}):
\begin{equation}
\begin{split}
&\epsilon_lp_\phi+\epsilon_l'p_\phi'=0,\\
&\pm \epsilon_l \sqrt{1-\frac{p_\phi^2\cos^2\rho_v}{\gamma^2}}\pm \epsilon_l' \sqrt{1-\frac{p_\phi'^2\cos^2\rho_v}{\gamma^2}}+\epsilon_p=0.
\end{split}
\end{equation}
Due to the presence of a turning point in the effective potential for the radial motion, one is able to make the mass in the leg, attached to the horizon, negative. Furthermore, as in the section \ref{Con_def}, geodesics may approach the vertex both from inward and from outward depending on the relation between masses of particles, propagating along geodesics. The behaviour of the geodesic configuration is qualitatively and quantitatively the same as in that case and is described in the Fig. \ref{fig:vertices_4pt} and between equations (\ref{eq:mom_cons}) and (\ref{eq:vertex_def}). So, analogously to (\ref{eq:vertex_def}), one obtains the radial coordinate of the vertex $\rho_v$ as a function of the angular momentum $p_\phi$, conserved along one of the legs:
\begin{equation}
\frac{(\epsilon_lp_\phi)^2}{\gamma^2}\cos^2\rho_v= \mu^2,   
\end{equation}
where the parameter $\mu$ is the same as in the previous subsection. Analogously to (\ref{eq:dphi12}), one finds the angle between end points of external legs $\Delta \phi$ as a function of the conserved angular momentum $p_\phi$ for configurations $--$ ($\epsilon_l^2-\epsilon_l'^2\le \epsilon_p^2$) and $-+$ ($\epsilon_l^2-\epsilon_l'^2\ge \epsilon_p^2$) in the case of positive intermediate dimensions $\epsilon_p>0$: :
\begin{equation}
\label{eq:dt12}
\begin{split}
&\Delta \phi^{--}= \frac{1}{\gamma}\left[\log\frac{1+\gamma/|p_\phi|}{\sin \rho_v +\sqrt{\gamma^2/p_\phi^2-\cos^2\rho_v}}+\log\frac{1+\gamma/|p_\phi'|}{\sin \rho_v +\sqrt{\gamma^2/p_\phi'^2-\cos^2\rho_v}}\right],\\
&\Delta \phi^{-+}=\frac{1}{\gamma}\left[\log\frac{1+\gamma/|p_\phi|}{\sin \rho_v +\sqrt{\gamma^2/p_\phi^2-\cos^2\rho_v}}-\log\frac{1-\gamma/|p_\phi'|}{\sin \rho_v +\sqrt{\gamma^2/p_\phi'^2-\cos^2\rho_v}}\right].
\end{split}
\end{equation}
One also finds the angle $\Delta \phi$ for configurations $++$ ($\epsilon_l^2-\epsilon_l'^2\le \epsilon_p^2$) and $+-$ ($\epsilon_l^2-\epsilon_l'^2\ge \epsilon_p^2$) in the case of negative intermediate dimensions $\epsilon_p<0$:
\begin{equation}
\label{eq:dt34}
\begin{split}
&\Delta \phi^{++}= \frac{1}{\gamma}\left[-\log\frac{1-\gamma/|p_\phi|}{\sin \rho_v +\sqrt{\gamma^2/p_\phi^2-\cos^2\rho_v}}-\log\frac{1-\gamma/|p_\phi'|}{\sin \rho_v +\sqrt{\gamma^2/p_\phi'^2-\cos^2\rho_v}}\right],\\
&\Delta \phi^{+-}=\frac{1}{\gamma}\left[-\log\frac{1-\gamma/|p_\phi|}{\sin \rho_v +\sqrt{\gamma^2/p_\phi^2-\cos^2\rho_v}}+\log\frac{1+\gamma/|p_\phi'|}{\sin \rho_v +\sqrt{\gamma^2/p_\phi'^2-\cos^2\rho_v}}\right].
\end{split}
\end{equation}
The solution to all of the equations (\ref{eq:dt12}), (\ref{eq:dt34}) has a familiar form: 
\begin{equation}
\label{eq:vertex_through_angle_btz2}
\frac{1}{\cos\rho_v}=\frac{\epsilon_l+\epsilon_l'}{2\mu}\coth\theta\mp\frac{\sqrt{\epsilon_p^2+(\epsilon_l-\epsilon_l')^2\sinh^2\theta}}{2\mu\sinh\theta},
\end{equation}
where $\theta=\gamma \Delta \phi/2$. Choosing the minus sign, one obtains the solution to both equations in (\ref{eq:dt12}), and the choice of the plus sign gives the solution to both equations in (\ref{eq:dt34}). It follows from (\ref{eq:vertex_through_angle_btz2}) that $p_\phi>\gamma$ and $p_\phi'>\gamma$, so the solution we found is self-consistent. Solving the differential equation
\begin{equation}
\label{eq:BTZ_diff_eq2}
\frac{dS}{d\theta}=\frac{2\epsilon_lp_\phi}{\gamma}=\frac{2\mu}{\cos\rho_v},    
\end{equation}
we obtain the action of the whole geodesic configuration:
\begin{equation}
\label{eq:action_BTZ2}
S=(\epsilon_l+\epsilon_l')\log\sinh \theta+\epsilon_p\artanh\frac{\cosh \theta}{\sqrt{1+\beta^2\sinh^2\theta}}\mp(\epsilon_l-\epsilon_l')\log\left(|\beta|\cosh\theta+\sqrt{1+\beta^2\sinh^2\theta}\right),
\end{equation}
which is analogous to the formula (\ref{eq:action_defect}) and can be identified with the conformal block. So, considering geodesics at the constant time slice in the BTZ black hole background also allows one to reproduce the discontinuity in the semiclassical heavy-light 4-point conformal block on the sphere. Moreover, the qualitative behaviour of the extremum geodesic configuration with the change of $\epsilon_p$ at this slice is the same as for the conical defect, which is not a surprise from the point of view of global AdS${}_3$ (see Appendix \ref{AppGlobalAdS}).

\section{Conclusion}
In this paper we have analyzed some analytic properties of semiclassical conformal blocks. It was questioned what is the counterpart of the resummation procedure of the instanton contributions near the poles suggested in \cite{Beccaria,GorskyMS} from the point of view of semiclassical conformal blocks and geodesic networks. It was demonstrated that semiclassical conformal blocks change discontinuosly at integer values of the parameter $\nu$ (or at $\nu=1$ for the linearized limit), which defines the intermediate dimension in the semiclassical limit, due to branch cuts in the complex $\nu$-plane that cross the real axis. The same phenomenon takes place also for the expectation value of the operator $L_0$ over the Gaiotto state. 

The linearized semiclassical torus conformal block was calculated analytically with the use of its holographic interpretation as the length of the tadpole graph embedded in thermal AdS${}_3$. We have shown that this geodesic configuration changes qualitatively when moving from values $\nu<1$ to $\nu>1$ (or, equivalently, from positive values of the mass in the loop to negative ones) as a consequence of the momentum conservation law in the vertex. Thus, the linearized torus block changes discontinuously at $\nu=1$ in accordance with the analysis of its resummed series. 

Finally, an analogous analysis of the behavior of the configuration of geodesics was accomplished in the case of the semiclassical 4-point heavy-light conformal block on the sphere. We generalized the result \cite{HijanoKraus} to the case of negative intermediate dimensions which allowed us to see different limits of the heavy-light conformal block when $\nu\to 1\pm 0$. Similarly to the case of the torus conformal block, the configurations of geodesics in the background of the conical defect in AdS${}_3$, dual to the heavy-light conformal block, are different depending on the sign of $\epsilon_p$.     

There are many questions to be elaborated further. The most immediate one concerns the microscopic mechanism which provides the disappearance of the massless W-boson with non-vanishing angular momentum from the spectrum. In the undeformed case the curve of marginal stability (CMS) in the moduli space does the job, so that the W-boson decays into a dyon and a monopole and no longer
exists as a stable particle inside the CMS. This process can be seen in many ways, for instance in terms of geodesics \cite{lerche}
or  string junctions \cite{sethi}. As was mentioned in \cite{GorskyMS} there is a natural analogue of the 
CMS in the deformed case at large values of the deformation parameter since the twisted superpotential for the CP$(1)$ model emerges in this limit upon resummation.

However, it is unclear what is the analogue of the decay of the W-boson into the monopole and the dyon at the CMS in the $\Omega$-deformed case. First, note that 
the spectrum of BPS states in the $\Omega$-deformed theory is more rich \cite{ito,hellerman,bcgk}. Namely, apart from BPS particles
there are extended objects in the spectrum, such as strings and domain walls, whose tensions are proportional to the deformation parameter and which are heavy and semiclassical in the limit of large deformation parameter. The immediate 
inspection of BPS masses of W-bosons and dyons shows that there is no naive  collinearity of masses at the CMS in the deformed case. Hence, a more complicated pattern of the decay of the W-boson with non-vanishing angular momentum should happen, which 
probably involves the extended states. Certainly, this point deserves further investigation.

Another point concerns a generalization of the analysis to the case when the defect is inserted into the deformed SYM theory and the quantum integrable system emerges. It was argued in \cite{GorskyMS} that the resummation
of instantons near the poles corresponds to the opening of tiny forbidden zones well above the barrier in the spectrum of the corresponding Hamiltonian. According to the AGT duality, the defect brings an additional operator inside the conformal block. It is interesting to understand if the insertion of the additional operator amounts to the same interpretation of the jump of the
order parameter as the rearrangement of the geodesic configuration. In other words, we question how the
non-perturbative phenomena in the quantum mechanics can be linked with the dynamics of particles in AdS${}_3$.

One more question concerns the transition from the NS limit to the  
limit of two small deformation parameters which are eventually switched off. All the CMS's around the higher poles at $a=n\epsilon_1$ should
glue into the single CMS around $a=0$ in the Seiberg-Witten solution.

Our work was supported
by RFBR grant 19-02-00214. 
The work of A.G. and M.L. was supported by Basis Foundation fellowship. A.G. thanks Simons Center for Geometry and Physics at Stony Brook University and Kavli Institute at UCSB where the part of work has been done for the hospitality and support.

\newpage
\appendix
\numberwithin{equation}{section}

\section{Global AdS${}_3$}
\label{AppGlobalAdS}
In this appendix we make a  comment on how geodesic configurations considered in sections \ref{Con_def} and \ref{BTZ_const_time} relate to each other from the point of view of global AdS${}_3$, where both geometries may be embedded into. 
\par To obtain the embedding of the BTZ black hole into the global AdS${}_3$, it is convenient to start from the embedding of AdS${}_3$ into the 4-dimensional space, covered by coordinates $(X_1,X_2,T_1,T_2)$, with the flat metric:
\begin{equation}
ds^2=-dT_1^2-dT_2^2+dX_1^2+dX_2^2.    
\end{equation}
Thermal AdS${}_3$ is a Riemannian manifold, defined by the following equation, with the metric induced from the flat space:
\begin{equation}
T_1^2+T_2^2-X_1^2-X_2^2=1.    
\end{equation}
The last equation is satisfied for the following embedding of a 3-manifold, covered by coordinates $(t,\rho,\phi)$: 
\begin{equation}
T_1=\frac{\cos t}{\cos \rho},\ T_2=\frac{\sin t}{\cos \rho},\ X_1=\tan \rho \cos \phi,\ X_2=\tan \rho \sin \phi.
\end{equation}
The induced metric on that manifold is that of AdS${}_3$ space:
\begin{equation}
ds^2=\frac{1}{\cos^2 \rho}\left(-dt^2+d\rho^2+\sin^2\rho\, d\phi^2\right), 
\end{equation}
where $0\le\rho<\pi/2$, $\phi\sim \phi+2\pi$ and we take $-\infty<t<\infty$, considering thus the universal covering of the initial space. The outer region of the BTZ black hole is embedded into the thermal AdS${}_3$ as follows:
\begin{equation}
X_1=\frac{ \sinh\left(\gamma \phi\,\right)}{\cos \rho},\ T_1=\frac{\cosh\left(\gamma \phi\,\right)}{\cos\rho},\ X_2=\tan \rho\cosh\left(\gamma t\right),\ T_2=\tan \rho\sinh\left(\gamma t\right),
\end{equation}
where $-\infty<t<+\infty,\ 0\le\rho<\pi/2,\ \phi\sim \phi+2\pi$. The induced metric on that manifold is that of the outer region of the BTZ black hole. So, putting tildes over AdS${}_3$ coordinates,
one obtains that the outer region of the BTZ black hole is embedded into the global AdS${}_3$ as follows:
\begin{equation}
\label{eq:BTZ_embedding}
\begin{split}
&\tan \tilde\rho \cos\tilde\phi =\frac{ \sinh\left(\gamma \phi\,\right)}{\cos \rho},\\
&\tan \tilde\rho \sin \tilde\phi = \tan \rho\cosh\left(\gamma t\right),\\
&\frac{\cos\tilde t}{\cos \tilde\rho}=\frac{\cosh\left(\gamma \phi\,\right)}{\cos\rho},\\
&\frac{\sin \tilde t}{\cos \tilde \rho}=\tan \rho\sinh\left(\gamma t\right).
\end{split}
\end{equation}
From these equations one sees that the BTZ constant-time slice $t=0$ maps into the AdS${}_3$ constant-time slice $\tilde t=0$, and the line $t=0,\ \phi=0$ maps into the line $\tilde t=0,\ \tilde \phi= \pi/2$. It explains why in subsection \ref{BTZ_const_time} we required the geodesic with the mass $\epsilon_p$ to satisfy the equation $d\phi=0$. Such geodesic configurations in the BTZ black hole background are in one-to-one correspondence with those geodesic configurations in AdS${}_3$, for which the equation $\tilde \phi =\pi/2$ holds for the geodesic with the mass $\epsilon_p$.

\section{BTZ black hole. Geodesics at constant angle}
\label{AppBTZ_const_angle}

In this appendix we demonstrate how the semiclassical heavy-light 4-point conformal block on the sphere is reproduced holographically as the length of geodesics at a constant angle in the BTZ black hole background. These geodesics are obtained from those in the conical defect geometry by analytic continuation $\alpha=i\gamma$. However, we are not able to reproduce the conformal block with a negative intermediate dimension in this way and thus to interpret the conformal block discontinuity in terms of the behaviour of geodesics at a constant angle in the BTZ geometry. 

We consider geodesic configurations in the space with the metric (\ref{eq:BTZ_metric_2}). We do not rename coordinates in this section intentionally, so that every equation below is obtained from a corresponding equation from section \ref{Con_def} only by the substitution $\alpha=i\gamma$. Separate geodesics, considered in the conical defect section, satisfied the following set of equations: $dt=0,\ u^2=1,\ p_\phi=\operatorname{const}$. Imposing the same constraints on geodesics in the BTZ black hole background, we thus consider non-rotating particles, moving along space-like trajectories, for which the energy conservation law holds true:
\begin{align}
&1.\, u^2=1 \Leftrightarrow -\gamma^2 \tan^2 \rho\,  \dot\phi^2+\frac{1}{\cos^2\rho}\,\dot \rho^2=1,\\ 
&2.\, p_\phi=\operatorname{const} \Leftrightarrow \dot \phi=\frac{p_\phi}{\gamma^2\tan^2\rho},
\end{align}
where derivatives are taken with respect to the proper length along geodesics. From these equations, one obtains the effective potential for the radial motion without turning points (compare to (\ref{eq:defect_eff_potential})):
\begin{figure}
\centering
\includegraphics[scale=0.22]{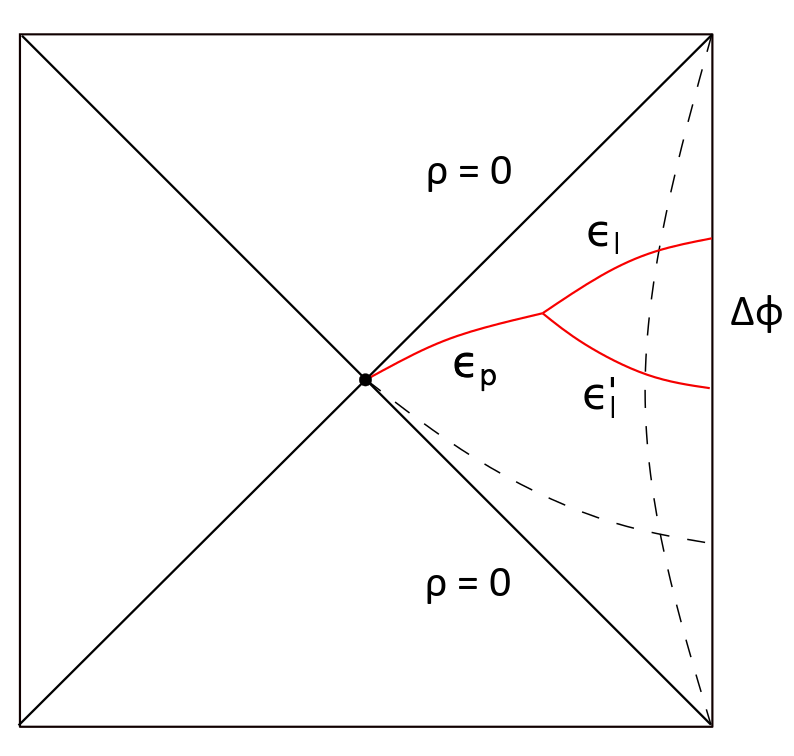}
\caption{The configuration of geodesics in the background of the BTZ black hole at the constant angle slice, which is dual to the semiclassical 4-point heavy-light conformal block on the sphere. The right triangle of this Penrose diagram is described by the metric (\ref{eq:BTZ_metric_1}) at $t=\operatorname{const}$. The dashed lines are slices of surfaces $\phi=\operatorname{const}$ and $\rho=\operatorname{const}$. The horizon is depicted by diagonal solid lines.}
\label{fig:BTZ}
\end{figure}
\begin{equation}
\label{eq:btz_eff_potential}
\left(\frac{d\rho}{d\phi}\right)^2+V(r)\equiv\left(\frac{d\rho}{d\phi}\right)^2+\gamma^2\sin^2 \rho \left(-\frac{\gamma^2 \tan^2 \rho}{p_\phi^2}-1\right)=0, 
\end{equation}
which means that one will not be able to make the mass in the leg, corresponding to the intermediate dimension, negative without violating the momentum conservation law in the vertex if one considers only geodesic configurations at the slice $t=\operatorname{const}$ restricted to the exterior region of the BTZ black hole. The momentum conservation law in the vertex in the black hole case coincides again with what we would obtain from (\ref{eq:mom_cons}) by the analytic continuation:
\begin{equation}
\label{eq:mom_cons_BTZ1}
\begin{split}
&\epsilon_lp_\phi+\epsilon_l'p_\phi'=0,\\
&- \epsilon_l \sqrt{1+\frac{p_\phi^2}{\gamma^2 \tan^2\rho_v}}- \epsilon_l' \sqrt{1+\frac{p_\phi'^2}{\gamma^2 \tan^2\rho_v}}+\epsilon_p=0,
\end{split}    
\end{equation}
except that now we have only one option for choosing sings in front of square roots. In the first equation we supposed that for the leg attached to the horizon $p_\phi=0$ in the vertex, which will be explained below. These equations have a real solution for the vertex radial position only when $\epsilon_p\ge \epsilon_l+\epsilon_l'$:
\begin{equation}
\frac{(\epsilon_lp_\phi)^2}{\gamma^2 \tan^2 \rho_v}= -\mu^2,     
\end{equation} 
where the definition of the parameter $\mu^2$ is given in the eqn. (\ref{eq:vertex_def}). The final question, which is left, is where exactly the geodesic, corresponding to the intermediate dimension, should be attached to the horizon for the action of the geodesic configuration to reproduce the conformal block. The equations $t=\operatorname{const},\ \rho=0$ define a one-dimensional surface rather than a single point. In the conical defect case that geodesic solved the equation $d\phi=0$, so in the black hole case we also impose the same equation on that geodesic, despite of the fact that it is not a necessary condition for geodesics attached to the horizon. Thus, the trajectory of a particle with the mass $\epsilon_p$ is attached to the "center"\ of the Kruskal-like extended BTZ geometry (see Fig. \ref{fig:BTZ}). From the eqn. (\ref{eq:btz_eff_potential}), one obtains the time separation $\Delta \phi$ between boundary points of geodesics, corresponding to "light"\ external dimensions, as a function of the conserved energy $p_\phi$:
\begin{equation}
\label{eq:BTZ_Delta_phi_const_angle}
\Delta \phi =\frac{1}{\gamma}\left[\log\frac{1/\sin\rho_v+\sqrt{\cot^2\rho_v+\gamma^2/p_\phi^2}}{1+\gamma/|p_\phi|}+ \log\frac{1/\sin\rho_v+\sqrt{\cot^2\rho_v+\gamma^2/p_\phi'^2}}{1+\gamma/|p_\phi'|}\right],    
\end{equation}
from where we find the radial position of the vertex:
\begin{equation}
\tan \rho_v=-\frac{\epsilon_l+\epsilon_l'}{2\sqrt{-\mu^2}}\coth \tilde \theta+\frac{\sqrt{\epsilon_p^2+(\epsilon_l-\epsilon_l')^2\sinh^2\tilde \theta}}{2\sqrt{-\mu^2}\sinh\tilde\theta},
\end{equation}
where $\tilde \theta\equiv \gamma\Delta\phi/2$. That solution is positive for small enough values of the parameter $\tilde \theta$. Let us also note that both of the last two equations can be obtained from (\ref{eq:dphi12}) and (\ref{eq:vertex_pos}) by the analytic continuation. Finally, we are able to find the action of the whole geodesic configuration, integrating the following equation:
\begin{equation}
\frac{dS}{d\tilde\theta}=2\sqrt{-\mu^2} \tan \rho_v. 
\end{equation}
Thus, the action is given by the following formula (note that we consider the case $\epsilon_p>0,\ \epsilon_l\ge\epsilon_l'$):
\begin{equation}
\label{eq:action_BTZ1}
S=-(\epsilon_l+\epsilon_l')\log\sinh \tilde\theta-\epsilon_p\artanh\frac{\cosh \tilde\theta}{\sqrt{1+\beta^2\sinh^2\tilde\theta}}+(\epsilon_l-\epsilon_l')\log\left(\beta\cosh\tilde\theta+\sqrt{1+\beta^2\sinh^2\tilde\theta}\right),
\end{equation}
which can be also obtained from (\ref{eq:action_defect}) by the substitution $\alpha=i\gamma$. Concluding this subsection let us emphasize its main points. First, the 4-point conformal block on the sphere with conformal dimensions, satisfying the condition $\epsilon_p\ge\epsilon_l+\epsilon_l'$, is reproduced holographically as the action of the geodesic configuration in the BTZ black hole background, with the trajectory of the particle with the mass $\epsilon_p$ being attached to the "center"\ of the Kruskal-like extended geometry and with all the geodesics being located outside the horizon. Furthermore, all the equations describing that geodesic configuration are obtained from corresponding equations in the case of the conical defect by the analytic continuation. Second, the form of the effective potential (\ref{eq:btz_eff_potential}), namely the absence of reflection points, does not allow one to reproduce holographically the conformal block with a negative intermediate dimension, keeping external dimensions fixed and positive, and thus to see the change in the holographic picture when moving from $\epsilon_p>0$ to $\epsilon_p<0$.

\printbibliography

\end{document}